\documentclass[aps,prx,twocolumn,superscriptaddress]{revtex4-2} 

\usepackage{amsmath,amssymb}
\usepackage[margin=1.0in]{geometry}
\usepackage{graphicx}

\usepackage{subfigure}
\usepackage{float}

\usepackage[normalem]{ulem}

\usepackage{xcolor} 
\usepackage{soul}
\definecolor{myyellow}{RGB}{255,255,0}
\sethlcolor{myyellow}

\usepackage{titlesec}

\titleformat{\section}
  {\normalfont\large\bfseries}
  {\thesection}              
  {1em}                       
  {}     

\titleformat{\subsection}
  {\normalfont\bfseries}
  {\thesubsection}        
  {1em}                    
  {}     
\renewcommand{\thesection}{\arabic{section}}
\renewcommand{\thesubsection}{\thesection.\arabic{subsection}}

\usepackage[colorlinks=true,linkcolor=magenta,urlcolor=blue,citecolor=blue]{hyperref}

\begin{document}

\title{Hyperuniform Disorder in Photonic Crystal Slabs with Intrinsic non-Hermiticity} 

\author{Zeyu Zhang}
\affiliation{Department of Physics, The Pennsylvania State University, University Park, PA, USA}

\author{Koorosh Sadri}
\affiliation{Department of Physics, The Pennsylvania State University, University Park, PA, USA}

\author{Brian Gould}
\affiliation{Department of Physics, The Pennsylvania State University, University Park, PA, USA}

\author{Mikael C. Rechtsman}
\affiliation{Department of Physics, The Pennsylvania State University, University Park, PA, USA}

\begin{abstract}

Hyperuniform disorder is a type of correlated disorder characterized by vanishing spectral density at small wavevectors, making the configuration effectively homogeneous on long length scales. In photonics, hyperuniform disorder is promising for generating isotropic photonic pseudogaps and engineering photonic crystal waveguides. However, these studies are largely restricted to idealized lossless settings, although all photonic systems necessarily have loss. In this work, light propagation in photonic crystal slabs with imposed hyperuniform disorder is investigated theoretically and numerically. The system is intrinsically non-Hermitian due to radiative loss, with non-Hermiticity appearing as a complex effective mass of a quadratic photonic band. A theoretical framework for disorder scattering is analytically derived in Hermitian and non-Hermitian quadratic bands with real and complex effective mass, respectively. In contrast to the power law behavior $|\mathbf{k}|^\alpha$ observed in the Hermitian case (where $\alpha$ is the hyperuniformity exponent), the scattering loss in the non-Hermitian band is given by $C_0+C_{\beta_2}\cdot|\mathbf{k}|^{\beta_2}$, where $C_0$ is a finite constant and the exponent $\beta_2\leq 2$. Our theoretical predictions are verified with tight-binding and Finite-Difference Time-Domain simulations with realistic photonic crystal parameters, based on recent experiments.\\

\end{abstract}

\maketitle

\section*{1\quad Introduction}

In photonic systems, disorder that is present in the system is often spatially correlated, and cannot be thought of as simply spatial `white noise'. An interesting and widely studied example is {\it hyperuniform disorder}, a class of correlated disordered structures in which long-wavelength density fluctuations are strongly suppressed, making them effectively homogeneous at large length scales  \cite{torquato2003local, torquato2018hyperuniform}. Mathematically, the spectral density associated with the disorder is characterized by a power law $\tilde{\rho}(\mathbf{q}) \to q^{\alpha}$ (where $\alpha>0$ is the hyperuniformity exponent) as the Fourier component $q$ approaches zero, $q \to 0$ (where $q=|\mathbf{q}|$). This distinctive correlation property places hyperuniform systems in a unique regime between crystalline order and uncorrelated randomness, leading to fundamentally different behavior in disorder-induced scattering processes \cite{leseur2016high, kim2023effective, klatt2025transparency}.

In photonic materials, hyperuniform disorder has attracted growing interest and has been widely investigated for its ability to support isotropic photonic pseudogaps and realize novel device functionalities. Extensive recent work has demonstrated that appropriately designed hyperuniform dielectric structures can support complete and isotropic photonic pseudogaps despite the absence of periodicity, highlighting hyperuniformity as an alternative route to bandgap formation beyond conventional Bragg scattering \cite{florescu2009designer, man2013isotropic, muller2013silicon, froufe2016role, gkantzounis2017hyperuniform, klatt2022wave, siedentop2024stealthy, casiulis2025gyromorphs}. Apart from band gaps, hyperuniform disorder has been shown to profoundly alter wave transport properties, giving rise to reduced scattering, diffusion, and disorder-induced localization phenomena. This provides fertile ground for exploring fundamental wave transport physics \cite{froufe2017band, aubry2020experimental, monsarrat2021pseudo, granchi2022near, cheron2022wave, sgrignuoli2022subdiffusive, vynck2023light, kim2023theoretical}.  However, these studies on the wave dynamics of hyperuniform disorder focus on the Hermitian setting, completely neglecting the effects of loss in photonic devices, which is ever-present.  This leaves a clear knowledge gap on the properties of scattering in realistic hyperuniform systems. More specifically, in Hermitian (lossless) systems, scattering from weak hyperuniform disorder is dominated by single-scattering events, which only allow for scattering to modes of the same energy. In contrast, for non-Hermitian systems, power conservation is broken by non-Hermiticity, so the wave dynamics can be significantly altered by scattering between modes of different energy. 

Due to their non-Hermitian nature, large system size, and the ability to precisely tailor the structure, photonic crystal slabs are an ideal platform for investigating the interplay between hyperuniform disorder and non-Hermiticity. Since photonic crystal Bloch modes may leak out of the plane of the slab, the system is intrinsically non-Hermitian even in the absence of material absorption \cite{johnson1999guided,fan2002analysis,zhen2015spawning,lalanne2018light}. In addition, photonic crystal slabs can be realized in experiments with a large system size, reaching well beyond $\sim 10^6$ unit cells in two dimensions, leaving sufficiently large area for disorder and loss to have significant dynamical interplay \cite{zhang2025airy}. Moreover, advances in nanophotonic fabrication technology provide a tailorable control of local feature size in a photonic crystal slab at each site, allowing the implementation of designed hyperuniform disorder configurations with high precision \cite{barsukova2025stealthy}. The presence of an underlying band structure in a photonic crystal slab provides a powerful experimental observable, namely the linewidth associated with the photonic band, for quantifying the effects of scattering in the system. 

In this work, we employ a similar setting to our recent experimental work \cite{barsukova2025stealthy} to introduce hyperuniform disorder as a perturbation on top of a lattice that exhibits a quadratic band in its band structure. The quadratic band is non-Hermitian with a complex effective mass, such that both the frequency and the out-of-plane radiation loss are proportional to $k^2$ (full details may be found in Ref. \cite{barsukova2025stealthy}). We analytically derive the general theoretical framework for the scattering loss both in a Hermitian quadratic band (real effective mass) and in a non-Hermitian quadratic band (complex effective mass), and verify the theoretical results with tight-binding (TB) simulations and full-wave Finite-Difference Time-Domain (FDTD) simulations in a realistic photonic crystal setting. For hyperuniform disorder with hyperuniformity exponent $\alpha$ (i.e., the spectral density of the disorder goes like $\tilde{\rho}(\mathbf{q})\to q^\alpha$ for small $q$), the scattering loss in the Hermitian case is proportional to $k^\alpha$ at leading order. In the non-Hermitian case, however, the scattering loss is always a finite constant $C_0$ at leading order (in $k$), and the next leading order is of the form of $C_{\beta_2} \cdot k^{\beta_2}$, where the exponent $\beta_2$ never exceeds $\beta_2 = 2$. This may be considered a somewhat surprising result: even an arbitrarily small non-Hermitian component gives rise to a drastic change to the exponent of scattering with momentum. These results serve as a benchmark for the interaction between hyperuniform disorder and non-Hermiticity, providing insights for understanding the wave dynamics in non-Hermitian hyperuniform systems, as well as in device design.

\section*{2\quad Results and Discussions}

\subsection*{2.1\quad The non-Hermitian nature of photonic crystal slabs}

\begin{figure*}[ht]
  \begin{center}
    \includegraphics[width=0.9\textwidth]{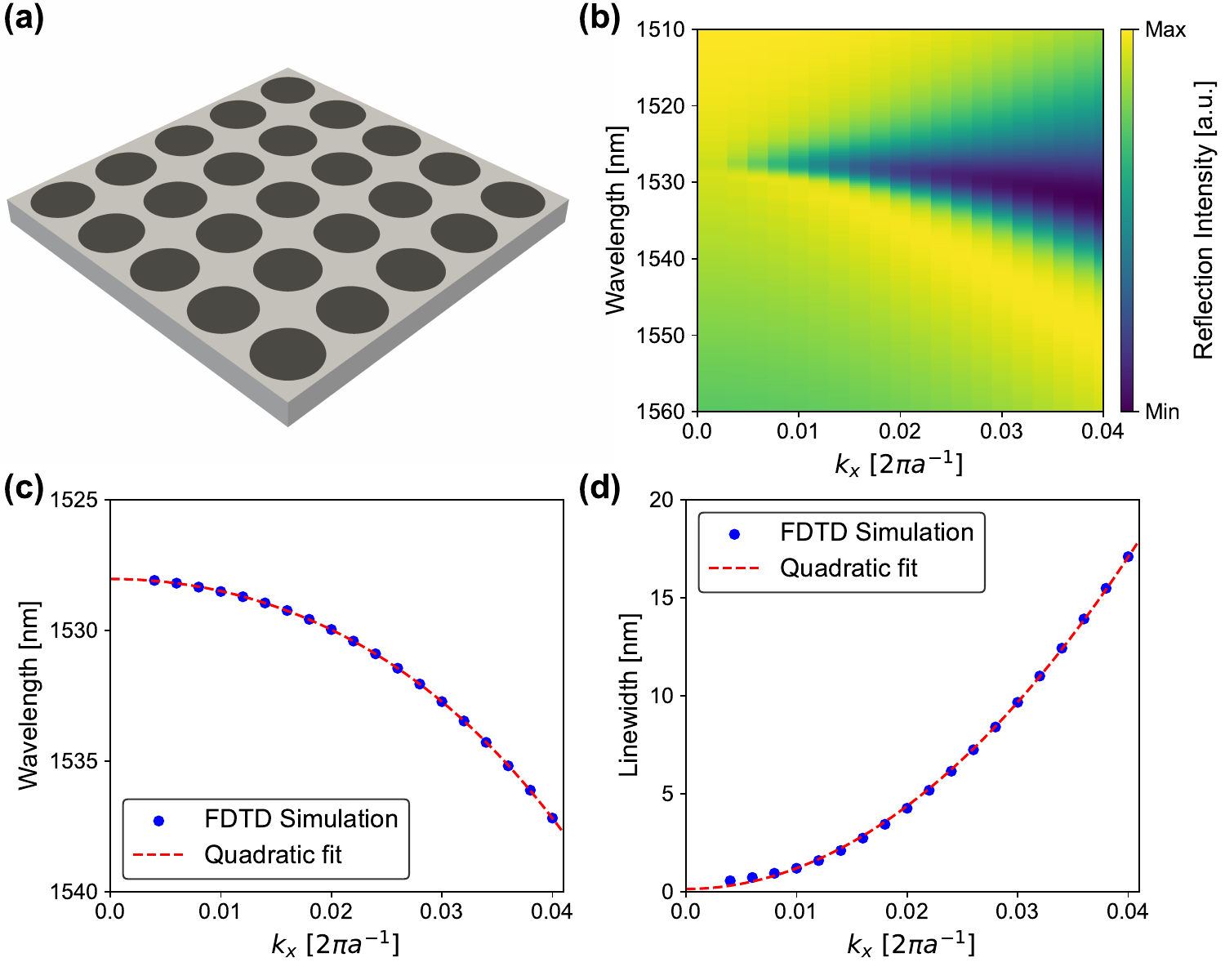}
    \caption{The non-Hermitian nature of photonic crystal slabs. (a) An illustration of the photonic crystal slab. The in-plane geometry contains circular air holes in a square lattice. (b) The simulated reflection spectrum of the structure in (a) along $k_y=0$. (c) The resonance frequency of the band extracted from (b). A quadratic fit is performed to obtain the real part of effective mass $\mathrm{Re}\left(\frac{1}{2m}\right)$. (d) The linewidth of the band extracted from (b) by fitting the reflection spectrum with a resonance lineshape. The linewidth also grows as $O(k^2)$, indicating the effective mass is complex.} 
    \label{figure1}
  \end{center}
\end{figure*}

We start from a periodic photonic crystal slab which hosts an isolated quadratic band in the transverse electric (TE)-like polarization. Figure \ref{figure1}(a) shows an illustration of the photonic crystal slab used in this work. The slab is made of silicon ($\varepsilon=12.11$) and the in-plane geometry consists of circular air holes ($\varepsilon=1$) arranged in a square lattice with lattice constant $a=1290nm$. The slab has a finite thickness $h=0.05a$ in the $z$ direction.

The eigenmodes of a photonic crystal slab are governed by Maxwell's equations, which lead to the following eigenvalue problem:
\begin{equation}
\mathbf{\nabla}\times \left[\varepsilon(\mathbf{r})^{-1}\mathbf{\nabla}\times\mathbf{H_k}\right]=E_\mathbf{k} \mathbf{H_k},
\label{equation1}
\end{equation}
where $\varepsilon(\mathbf{r})$ is the dielectric function, with $\mathbf{r}=(x,y,z)$ the position, $\mathbf{k}=\left(k_x, k_y\right)$ is the 2D lattice Bloch momentum, $\mathbf{H_k}$ is the magnetic field profile of the eigenmode, and $E_\mathbf{k}=\left(\omega_\mathbf{k}/c\right)^2$ is the energy of the eigenmode, with $\omega_\mathbf{k}$ the eigenfrequency and $c$ the speed of light. 

Two-dimensional photonic crystal slabs support both confined guided modes (which appear below the light line ($\omega\leq c k$) in the band structure) and leaky resonances that can couple to the radiation continuum \cite{fan2002analysis,photonicbook} (above the light line). In this work, we focus on radiative modes that are above the light line, centered at $\mathbf{k}=\Gamma=(0,0)$. Due to their radiative coupling, these modes have finite lifetimes, which can be incorporated into the formalism through a complex eigenvalue $E_\mathbf{k}$. The imaginary component $\mathrm{Im}\left(E_\mathbf{k}\right)$ quantifies the rate of out-of-plane radiative loss and is directly related to the spectral linewidth by $\Delta\lambda=-\frac{2\pi \mathrm{Im}\left(E_k\right)}{\mathrm{Re}\left(E_k\right)^\frac{3}{2}}$. More details can be found in Supplemental Information Section 1. Throughout this paper, the imaginary part of the eigenvalue $\mathrm{Im}\left(E_\mathbf{k}\right)$ in the periodic lattice is referred to as the intrinsic loss (or intrinsic linewidth), to distinguish it from the additional broadening induced by disorder scattering (or excess linewidth). For notation simplicity and analytical convenience, we describe band dispersions in terms of the eigenvalue 
$E=\left(\omega/c\right)^2$ of Maxwell's equations rather than the frequency $\omega$ itself. By analogy with the dispersion relation of a massive particle in quantum mechanics, this quantity is referred to as `energy', despite having a unit of $[a^{-2}]$.

The simulated reflection spectrum of the photonic crystal slab in the vicinity of the quadratic band is shown in Figure \ref{figure1}(b). The simulation results are obtained from \textsc{Tidy3D}, using the Finite-Difference Time-Domain (FDTD) method \cite{Minkov:24}. In the simulation, the photonic crystal slab is illuminated by a plane wave at a range of angles, and the reflection intensities at different frequencies are collected. An isolated quadratic band is shown in the band structure by a clear resonance lineshape at each Bloch momentum $k_x$.

The quadratic band in Figure \ref{figure1}(b) can be described as:
\begin{equation}
E_{\mathbf{k}}=E_0-\frac{1}{2m}\left(k_x^2+k_y^2\right),
\label{equation2} 
\end{equation}
where $E_\mathbf{k}$ is the energy of the band at wavevector $\mathbf{k}=\left(k_x,k_y\right)$ and $m$ is the effective mass. To determine the tip energy $E_0$ and the effective mass $m$ of the quadratic band shown in Figure \ref{figure1}(b), the wavelength and the linewidth associated with the photonic band are quantitatively extracted from the resonance lineshape \cite{fan2002analysis, barsukova2025stealthy}, which are plotted in Figure \ref{figure1}(c)(d), respectively. In the simulation, the tip energy is $E_0=28.1a^{-2}$, corresponding to wavelength $\lambda_0=1529nm$; and the effective mass is $\frac{1}{2m}=4.27-4.67i$. Importantly, the effective mass is inherently complex. This feature arises because both the real part of the band dispersion (wavelength) and the imaginary part associated with radiative loss (linewidth) exhibit a quadratic dependence on the in-plane wavevector, a characteristic resulting from the symmetry-protected bound state in the continuum (BIC) \cite{hsu2016bound, koshelev2018asymmetric}. The emergence of a complex effective mass thus reflects the intrinsic non-Hermitian nature of photonic crystal slabs and establishes them as an ideal platform for exploring hyperuniform disorder in a non-Hermitian setting.

\subsection*{2.2\quad Hyperuniform disorder in a Hermitian quadratic band}

\begin{figure*}[ht]
  \begin{center}
    \includegraphics[width=0.9\textwidth]{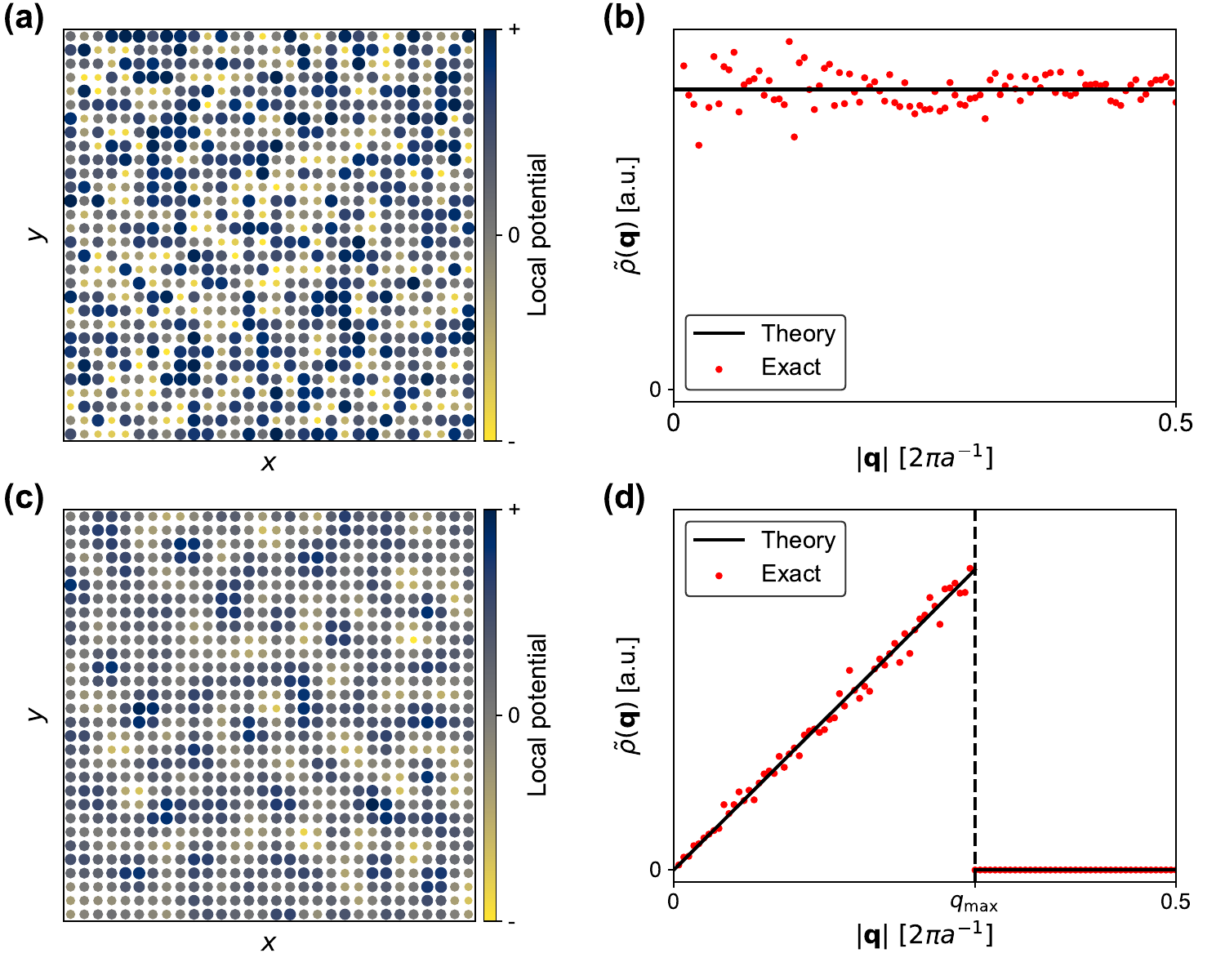}
    \caption{Hyperuniform disorder in photonic crystal slabs. (a) Local potential configuration with uncorrelated disorder. The color and the size of the holes represent the local potential change at each site with respect to the periodic (without disorder) case. Only an area of $30\times 30$ unit cells is shown. (b) The spectral density $\tilde{\rho}(\mathbf{q})$ corresponding to the disorder pattern in (a). The system size is $N_x\times N_y=500\times 500$. (c) Hyperuniform local potential configuration with hyperuniformity exponent $\alpha=1$. The potential is highly spatially correlated. (d) The spectral density $\tilde{\rho}(\mathbf{q})$ corresponding to the disorder pattern in (c). The spectral density scales as $\tilde{\rho}(\mathbf{q})\to O(k^1)$ when $k\to 0$. The Fourier components for $k>q_\mathrm{max}=0.3\ [2\pi a^{-1}]$ are also filtered out to make the spectral density $\tilde{\rho}(\mathbf{q})$ symmetric in all the directions.}
    \label{figure2}
  \end{center}
\end{figure*}

Next, we introduce a hyperuniform disorder pattern into the periodic photonic crystal slab. Figure~\ref{figure2} is a schematic diagram showing how we impose such disorder upon the crystal. Figure \ref{figure2}(a) demonstrates how random uncorrelated disorder is added to the photonic crystal slab. The color and size of the holes in the figure represent the change in radius. The disorder we impose can be treated as a random potential $V_{i,j}$ that varies in space, where $(i,j)$ denotes the site index. To establish a quantitative mapping between the variation in hole radius and the effective potential, we analyze the dependence of the band tip energy $E_0$ on the hole radius in a corresponding periodic structure. This relationship provides a direct calibration that allows us to quantitatively assign a value to the local potential \cite{zhang2025airy, barsukova2025stealthy}. More details can be found in Supplemental Information Section 2. In Figure \ref{figure2}(a), the local potential $V_{i,j}$ is randomly and uniformly selected from the range $\left[-w, w\right]$, where $w$ represents the degree of disorder, meaning an uncorrelated disorder configuration is generated.
 
Figure \ref{figure2}(b) shows the spectral density $\tilde{\rho}(\mathbf{q})$ of the uncorrelated disorder added in Figure \ref{figure2}(a). The spectral density is calculated from the Fourier transform of the potential configuration $\tilde{V}(\mathbf{q})$ by $\tilde{\rho}(\mathbf{q})=\left<\left|\tilde{V}(\mathbf{q})\right|^2\right>$, where $\left<\cdot\right>$ represents the average over configurations and $\mathbf{q}$ denotes the Fourier component. In the continuous limit (system size approaching infinity), the spectral density of an uncorrelated disorder is independent of the in-plane Fourier component $\mathbf{q}$, which is shown by a black line. However, due to the finite system size effect, the spectral density has small fluctuations around the targeted spectrum.

To generate a hyperuniform disorder configuration with hyperuniformity exponent $\alpha$, we perform a Fourier filtering method \cite{makse1996method} by setting $\tilde{V^\prime}(\mathbf{q})=\sqrt{\frac{\alpha+2}{2}}\left(\frac{q}{q_\mathrm{max}}\right)^{\frac{\alpha}{2}}\tilde{V}(\mathbf{q})$ when $q<q_\mathrm{max}$, and $\tilde{V^\prime}(\mathbf{q})=0$ for $q>q_\mathrm{max}$, where $q=|\mathbf{q}|$, and $q_\mathrm{max}=0.3\ [2\pi a^{-1}]$ is a cutoff wavenumber to make the spectral density isotropic in the first Brillouin zone. The hyperuniform configuration in real space $V^\prime_{i,j}$ is then generated by the inverse Fourier transform of $\tilde{V^\prime}(\mathbf{q})$. More details of the Fourier filtering method can be found in Supplemental Information Section 2. The corresponding spectral density added to the photonic crystal slab is of the form:
\begin{equation}
\tilde{\rho}(\mathbf{q})=
\left\{
\begin{aligned}
\frac{\alpha+2}{6}w^2 \left(\frac{q}{q_\mathrm{max}}\right)^\alpha & , & q<q_\mathrm{max}\\
0 & , & q>q_\mathrm{max}
\end{aligned}
\right.
\label{equation3}
\end{equation}

Importantly, for the spectral density in Equation \eqref{equation3}, the variance of the local potential in real space $\mathrm{Var}\left(V_{i,j}\right)=\frac{a^2 q_\mathrm{max}^2w^2}{12\pi}$ is independent of $\alpha$. This allows us to make fair comparisons across the results from different hyperuniformity exponents $\alpha$.

Figure \ref{figure2}(c) shows the disorder configuration in real space with $\alpha=1$. Compared to the uncorrelated disorder in Figure \ref{figure2}(a), the pattern of the disorder is highly spatially correlated. Figure \ref{figure2}(d) calculates the corresponding spectral density associated with Figure \ref{figure2}(c). Compared with the theoretical calculation obtained from Equation \eqref{equation3} (black line), the spectral density shows small fluctuations, again as a result of finite-size effects.

In a periodic lattice, the lattice Bloch momentum $\mathbf{k}$ is a conserved quantity, so the band structure is well defined. However, as the disorder breaks the periodicity, the lattice Bloch momentum $\mathbf{k}$ is no longer a good quantum number. Instead, the system is best described by the spectral response function $A_\mathbf{k}(E)$.  The spectral function inherits the properties of the Bloch modes, while broadening in a way that incorporates loss \cite{moroz1995density,fetter2012quantum,wiersma2013disordered}.  Specifically, in the weak disorder limit, the peak of the spectral function $A_\mathbf{k}(E)$ is still centered at $\mathrm{Re}(E_\mathbf{k})$ at leading order, but the full width at half maximum (FWHM) of the peak is broadened from intrinsic loss $-2\mathrm{Im}(E_\mathbf{k})$ to $-2\mathrm{Im}(E_\mathbf{k})-2\mathrm{Im}(\Sigma_\mathbf{k})$, where $\Sigma_\mathbf{k}$ is the self-energy, whose imaginary part represents addition scattering loss (excess linewidth) introduced by disorder. The self-energy can be calculated perturbatively in the Born approximation as:
\begin{equation}
\Sigma_\mathbf{k}=\frac{1}{N_x N_y}\sum_\mathbf{q} \frac{\tilde{\rho}(\mathbf{q})}{\mathrm{Re}\left(E_{\mathbf{k}}\right)-E_{\mathbf{k+q}}+i0^{+}}+O(\tilde{\rho}^2),
\label{equation4}
\end{equation}
where $N_x$ and $N_y$ are the system size in the $x$ and the $y$ directions, and $E_{\mathbf{k}}=E_0-\frac{k^2}{2m}$ is the quadratic band. 

In this section, we first consider the Hermitian case in which the tip energy $E_0$ and the effective mass $m$ are real, and leave the discussion of the effects of non-Hermiticity introduced by the imaginary part of the effective mass $\mathrm{Im}(m)$ to the next section. Given the spectral density in Equation \eqref{equation3} and the band dispersion relation in Equation  \eqref{equation2}, the self-energy in Equation \eqref{equation4} can be analytically calculated:
\begin{equation}
\mathrm{Im}\left(\Sigma_\mathbf{k}\right)
= 
-\frac{2^\alpha(\alpha+2)\cdot\Gamma\left(\frac{\alpha+1}{2}\right)\cdot w^2 a^2 m}{12\sqrt{\pi}\cdot\Gamma\left(\frac{\alpha+2}{2}\right)\cdot q_\mathrm{max}^\alpha}k^\alpha
+O(w^4),
\label{equation5}
\end{equation}
where $\Gamma$ is the Gamma function. More analytical derivations can be found in Supplemental Information Section 3.

\begin{figure*}[ht]
  \begin{center}
    \includegraphics[width=0.9\textwidth]{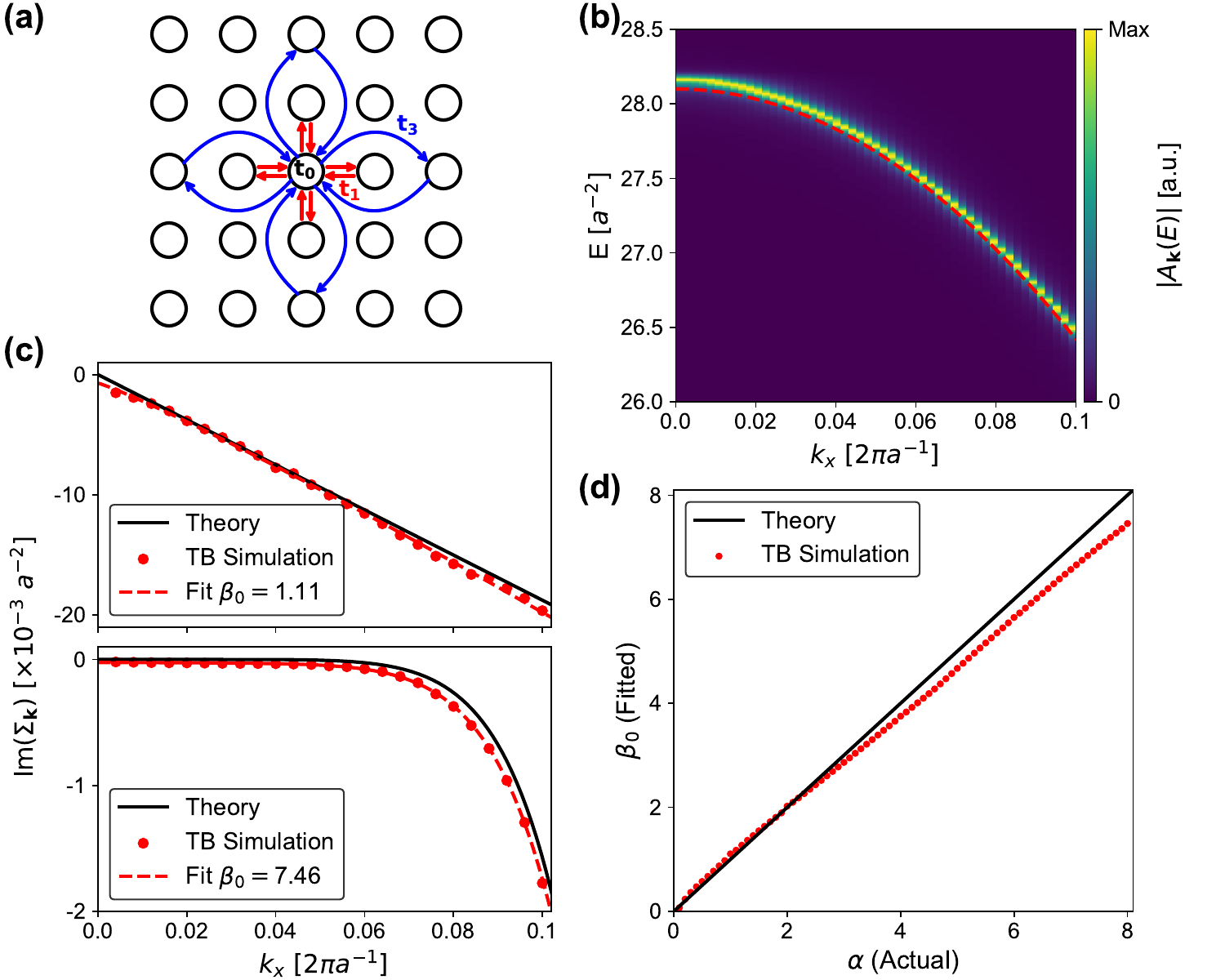}
    \caption{Effects of Hyperuniform disorder in a Hermitian quadratic band. (a) The tight-binding model used in the simulation. The on-site energy $t_0$, the nearest neighbor hopping $t_1$, and the third nearest neighbor hopping $t_3$ are considered. (b) The spectral function $A_\mathbf{k}(E)$ when an uncorrelated on-site disorder is imposed on the tight-binding model in (a) with a system size of $N_x\times N_y=500\times 500$. The red dashed line shows the unperturbed band structure. (c) The scattering loss $\mathrm{Im}\left(\Sigma_\mathbf{k}\right)$ along $k_y=0$ when a hyperuniform on-site disorder is imposed on the system. The simulation results for two different values of $\alpha$ are shown in the upper panel ($\alpha=1$) and the lower panel ($\alpha=8$). The scattering loss $\mathrm{Im}\left(\Sigma_\mathbf{k}\right)$ is fitted by $C_0+C_{\beta_0}\cdot k^{\beta_0}$ to obtain the power law behavior with respect to $k$. (d) The comparison between the actual hyperuniformity exponent $\alpha$ and the fitted exponent $\beta_0$.}
    \label{figure3}
  \end{center}
\end{figure*}

As shown in Equation \eqref{equation5}, the scattering loss $\mathrm{Im}\left(\Sigma_\mathbf{k}\right)$ at leading order (which only considers the single scattering) is proportional to $k^\alpha$. This is because both the group velocity and the radius of the iso-frequency contour grow as $O(k)$ in a quadratic band. Hence, the scattering loss $\mathrm{Im}\left(\Sigma_\mathbf{k}\right)$ has the same exponent $O(k^\alpha)$ as the spectral density. This is the key feature of hyperuniform wave dynamics in a Hermitian quadratic band. The theoretical result of Equation \eqref{equation5} is verified via a tight-binding (TB) simulation. Figure \ref{figure3}(a) illustrates the setup of the TB model. A square lattice with on-site energy $t_0$, nearest neighbor hopping $t_1$ and third nearest neighbor hopping $t_3$ is considered. In the TB simulation, the hopping parameters are set to be $t_0=E_0-\frac{5}{2m}a^{-2}$, $t_1=\frac{4}{3}\frac{1}{2m}a^{-2}$, and $t_3=-\frac{1}{12}\frac{1}{2m}a^{-2}$, such that the band dispersion near $k=0$ is isotropic and quadratic up to $O(k^6)$, as described in Equation \eqref{equation2}, with tip energy $E_0=28.1a^{-2}$ and effective mass $\frac{1}{2m}=4.27$.

Figure \ref{figure3}(b) shows the spectral function $A_\mathbf{k}(E)$ when an uncorrelated disorder ($\alpha=0$) is added to the system by a random on-site potential of strength $w=1.23a^{-2}$. The band structure in the periodic lattice without disorder is also shown with a red dashed line. The spectral function has a finite linewidth, even in the absence of intrinsic loss ($\mathrm{Im}(E_\mathbf{k})=0$). This is due to the scattering loss ($\mathrm{Im}(\Sigma_\mathbf{k})$) introduced by disorder. According to Equation \eqref{equation5}, $\mathrm{Im}(\Sigma_\mathbf{k})$ is a constant when $\alpha=0$, indicating that the scattering loss is independent of $k$. We also note that the peak of the spectral function is close to the band structure in the clean lattice (without disorder), which is consistent with our analysis earlier in the weak disorder regime. The small shift between them is caused by the real part of self-energy, as discussed in detail later.

The Fourier filtering method illustrated in Figure \ref{figure2} is used to generate hyperuniform disordered configurations with $\alpha > 0$. Figure \ref{figure3}(c) presents two typical examples for $\alpha = 1$ (upper panel) and $\alpha = 8$ (lower panel). The simulation results are extracted from the FWHM width of the spectral function, while the theoretical calculation is obtained from Equation \eqref{equation5} with no free parameters. We observe quantitative agreement between the two; however, the simulation results are consistently slightly larger than the theoretical predictions. This discrepancy is due to finite-size effects arising from the need for a non-zero $i0^+$ term in Equation \eqref{equation4}. To account for this, the simulation data is fitted using $\mathrm{Im}(\Sigma_\mathbf{k}) = C_0 + C_{\beta_0} \cdot k^{\beta_0}$ with fitting parameters $C_0$, $C_{\beta_0}$, and ${\beta_0}$. The fitted curves and the corresponding values of $\beta_0$ are also shown in Figure \ref{figure3}(c) as red dashed lines.

Figure \ref{figure3}(d) plots the fitting result of the exponent $\beta_0$ as a function of the hyperuniformity exponent $\alpha$. The simulation results are quantitatively consistent with the theoretical analysis. In extreme cases with large hyperuniformity exponent $\alpha$, the fitted exponent $\beta_0$ in the simulation data is slightly lower than the theoretical prediction. This is due to finite-size effects, as explained earlier. From the simulation results in Figure \ref{figure3} and the theoretical calculations of Equation \eqref{equation5}, the $k^\alpha$ exponent of scattering loss $\mathrm{Im}\left(\Sigma_\mathbf{k}\right)\propto k^\alpha$ is the key signature of the wave dynamics of hyperuniform disorder in a Hermitian quadratic band.

\subsection*{2.3\quad Hyperuniform disorder in a non-Hermitian quadratic band}

Next, we turn our attention to the wave dynamics of hyperuniform disorder in a non-Hermitian quadratic band. When the effective mass is complex, the formalism in Equation \eqref{equation4} is still valid for weak disorder, but the theoretical results become significantly more complex.  In the non-Hermitian case, the scattering loss is expressed as follows:
\begin{widetext}
\begin{equation}
\mathrm{Im}\left(\Sigma_\mathbf{k}\right)=\frac{(\alpha+2)w^2 a^2}{6\pi\alpha}\cdot\mathrm{Im}(m)+
w^2\times\left\{
\begin{aligned}
O(k^\alpha)
& , & 0<\alpha<\frac{2}{\pi}\arctan\left(\frac{\mathrm{Re}(m)}{\mathrm{Im}(m)}\right)+1\\
O(k^2)
& , & \alpha=\frac{2}{\pi}\arctan\left(\frac{\mathrm{Re}(m)}{\mathrm{Im}(m)}\right)+1\\
O(k^\alpha)
& , & \frac{2}{\pi}\arctan\left(\frac{\mathrm{Re}(m)}{\mathrm{Im}(m)}\right)+1<\alpha<2\\
O(k^2\ln(k))
& , & \alpha=2\\
O(k^2)
& , & \alpha>2
\end{aligned}
\right.
+O(w^4)
\label{equation6}
\end{equation}
\end{widetext}

Compared to the analogous Hermitian case of Equation \eqref{equation5}, non-Hermiticity significantly alters the wave dynamics in hyperuniform systems. Instead of a power law of $k^\alpha$ in the Hermitian case, the scattering loss $\mathrm{Im}\left(\Sigma_\mathbf{k}\right)$ in a non-Hermitian quadratic band is always a finite constant at leading order in $k$. This finite constant, $\frac{(\alpha+2)w^2 a^2}{6\pi\alpha}\cdot\mathrm{Im}(m)$, is proportional to the imaginary part of the effective mass $\mathrm{Im}(m)$, which clearly shows its non-Hermitian origin. When $\mathrm{Im}(m)=0$, Equation \eqref{equation6} reduces to the Hermitian case, as described by Equation \eqref{equation5}. More detailed derivations can be found in Supplemental Information Sections 3 and 4.

\begin{figure*}[ht]
  \begin{center}
    \includegraphics[width=0.9\textwidth]{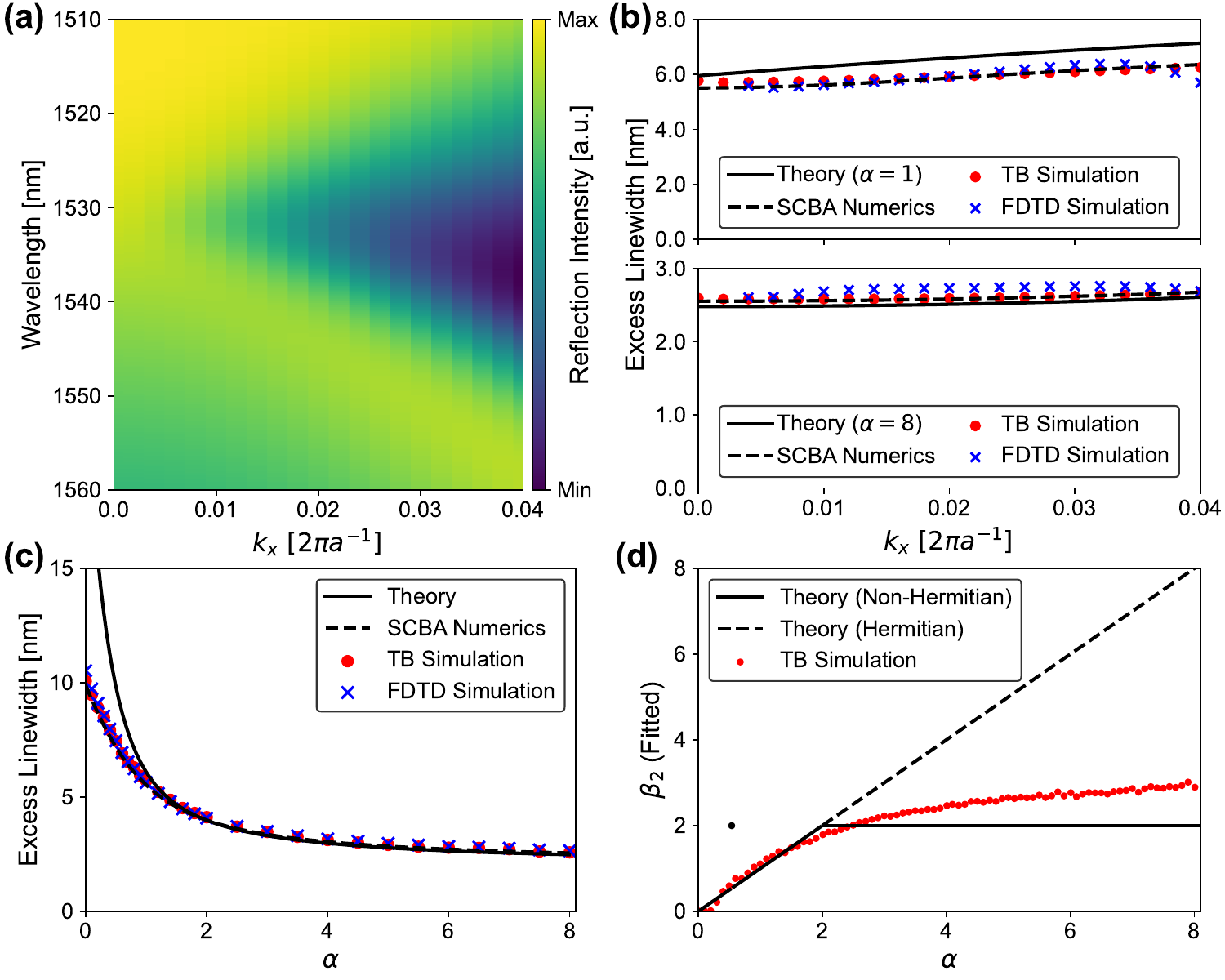}
    \caption{Effects of Hyperuniform disorder in a non-Hermitian quadratic band. (a) The simulated reflection spectrum of the same quadratic band as in Figure \ref{figure1}(b) when an uncorrelated disorder (with disorder strength $w=3.44a^{-2}$) is imposed on the photonic crystal slab. The band becomes blurry because the linewidth is broadened by scattering loss $\mathrm{Im}\left(\Sigma_k\right)$ introduced by disorder. The system size is $N_x\times N_y=100\times 100$. (b) The excess linewidth along $k_y=0$ when $\alpha=1$ (upper panel) and $\alpha=8$ (lower panel). The FDTD simulation is averaged over 10 disorder configurations each with system size $N_x\times N_y=100\times 100$. The TB simulations and SCBA numerics are obtained from a system size of $N_x\times N_y=500\times 500$. (c) The excess linewidth at $k_x=0.004\ [2\pi a^{-1}]$ and $k_y=0$ when $\alpha$ is swept. (d) The excess linewidth is fitted by $C_0+C_{\beta_2}\cdot k^{\beta_2}$ in the tight-binding simulation to obtain the exponent of the second leading order $\beta_2$.}
    \label{figure4}
  \end{center}
\end{figure*}

We verify the theoretical calculations of Equation \eqref{equation6} in TB and FDTD simulations that correspond to a realistic photonic setting. Figure \ref{figure4}(a) shows the FDTD simulated reflection spectrum when an uncorrelated disorder ($\alpha=0$) with $w=3.44a^{-2}$ is imposed on the same photonic crystal slab structure as in Figure \ref{figure1}. Compared with Figure \ref{figure1}(b), the band feature in the disordered system becomes ``blurry'' because the linewidth is broadened by the scattering loss $\mathrm{Im}\left(\Sigma_\mathbf{k}\right)$ introduced by disorder. The excess linewidth $\Delta \lambda$ is related to the scattering loss $\mathrm{Im}\left(\Sigma_k\right)$ by $\Delta\lambda=-\frac{2\pi \mathrm{Im}\left(\Sigma_k\right)}{\mathrm{Re}\left(E_k\right)^\frac{3}{2}}$, and can be obtained by the difference between the linewidth in a disordered sample (for example, linewidth in Figure \ref{figure4}(a)) and the intrinsic linewidth in the periodic sample (linewidth in Figure \ref{figure1}(b)).

Figure \ref{figure4}(b) simulates the excess linewidth $\Delta \lambda$ for the cases of $\alpha=1$ (upper panel) and $\alpha=8$ (lower panel) as a function of $k_x$ where $k_y=0$. In the TB simulation, the effective mass is set to be $\frac{1}{2m}=4.27-4.67i$, which is the same as the photonic band in the FDTD simulation in Figure \ref{figure1}. Note that the effective mass is complex, so the TB model is non-Hermitian; the complex effective mass is introduced into the TB model by using appropriate complex hoppings. The excess linewidth $\Delta \lambda$ in the FDTD simulation is obtained from the linewidth extracted from the reflection spectrum. In Figure \ref{figure4}(b), only the FDTD simulation data for $k_x\geq 0.004\ [2\pi a^{-1}]$ are shown. This is because the BIC pins the radiative linewidth to zero at $\mathbf{k}=\Gamma$, making it difficult to identify the band resonance feature or reliably extract the linewidth near $k_x=0$. 

The data in FDTD and TB simulations are highly consistent with the theoretical prediction of Equation \eqref{equation6} (solid black line) when $\alpha$ is large (for example, $\alpha=8$ in the lower panel of Figure \ref{figure4}(b)). However, when the hyperuniformity exponent $\alpha$ is small, as in the upper panel where $\alpha=1$, the FDTD simulation and the TB simulation deviate from the theoretical calculation with a lower excess linewidth $\Delta \lambda$. This is due to higher-order effects from multiple scattering. In the theoretical calculation of Equation \eqref{equation6}, only the leading-order single-scattering events ($O(w^2)$ terms) are considered, while higher-order effects from multiple scattering (the $O(w^4)$ term) are neglected. This approximation is accurate and valid when the self-energy $\Sigma_\mathbf{k}$ (both real part and imaginary part) is small, such that the disorder does not shift the band energy or linewidth at leading order.  This can be seen in Figure \ref{figure3}(b): the peak of the spectral function $A_\mathbf{k}(E)$ is approximately the same as the band energy in the periodic lattice. However, when $\alpha$ is small, the imaginary part of the self-energy $\mathrm{Im}\left(\Sigma_\mathbf{k}\right)$ calculated from Equation \eqref{equation6} (as well as the real part of the self-energy) is large and significantly shifts the peak of the spectral function $A_\mathbf{k}(E)$ and the linewidth of the underlying photonic band. The scattering loss $\mathrm{Im}(\Sigma_k)$ can no longer be evaluated at $E=\mathrm{Re}(E_\mathbf{k})$ with band dispersion $E_\mathbf{k+q}$, which is the case shown in Equation \eqref{equation4}, but needs to be modified according to:
\begin{equation}
\scalebox{0.8}{$\displaystyle
\Sigma_\mathbf{k}=\frac{1}{N_x N_y}\sum_\mathbf{q} 
\frac{\tilde{\rho}(\mathbf{q})}{\mathrm{Re}(E_{\mathbf{k}}+\Sigma_\mathbf{k})-\left(E_{\mathbf{k+q}}+\Sigma_\mathbf{k+q}\right)+i0^{+}}+O(\tilde{\rho}^2),
\label{equation7}
$}
\end{equation}
where the extra term $\mathrm{Re}(\Sigma_\mathbf{k})$ on the right hand side reflects the shift of the peak energy in the spectral function, and the extra $\Sigma_\mathbf{k+q}$ term represents the shift in photonic band dispersion.  Importantly, the self-energy ($\Sigma_\mathbf{k}$ or $\Sigma_\mathbf{k+q}$) is found on both sides of Equation \eqref{equation7}, and therefore must be solved for self-consistency: this is the self-consistent Born approximation (SCBA) \cite{lee1985disordered}. The SCBA numerical results are shown in Figure \ref{figure4}(b) with dashed lines. The deviation between theoretical analysis and simulation results is largely captured by the SCBA numerical calculations, confirming that the source of this discrepancy is from higher-order multiple scattering.  Clearly, the SCBA gives highly accurate results that capture multiple scattering; this comes at the cost of giving up the ability to derive analytical expressions, as in Equation \eqref{equation6}.

Next, we focus on the scattering loss $\mathrm{Im}(\Sigma_\mathbf{k})$ as a function of the hyperuniformity exponent, $\alpha$. According to Equation \eqref{equation6}, the scattering loss $\mathrm{Im}(\Sigma_\mathbf{k})$ is always a finite constant $\frac{(\alpha+2)w^2 a^2}{6\pi\alpha}\cdot\mathrm{Im}(m)$ at leading order, which is proportional to $\mathrm{Im}(\Sigma_\mathbf{k})\propto \frac{\alpha+2}{\alpha}$. This means that as the hyperuniformity exponent $\alpha$ decreases, the scattering loss $\mathrm{Im}(\Sigma_\mathbf{k})$ increases as $\frac{\alpha+2}{\alpha}$ and diverges when $\alpha\to 0$. Figure \ref{figure4}(c) shows the comparison of the excess linewidth $\Delta \lambda$ at $k_x=0.004\ [2\pi a^{-1}]$ across the theoretical analysis (black solid line, calculated from Equation \eqref{equation6}), TB simulations (red dots, obtained from the FWHM width of the spectral function $A_{\mathbf{k}}(E)$), and FDTD simulations (blue crosses, extracted from the linewidth associated with the photonic band). When $\alpha$ is large, the TB simulations and FDTD simulations are highly consistent with theoretical predictions. However, as $\alpha$ decreases to $\alpha<1$, both TB simulations and FDTD simulations clearly deviate from the theory. This is due to the higher-order effects arising from multiple scattering. We use the same SCBA numerical calculation described by Equation \eqref{equation7} to estimate the excess linewidth, and plot the results with a dashed black line. We again observe quantitative agreement between the simulation results (from both TB and FDTD) and the theoretical calculations.

Finally, we explore the next leading-order effect in $k$ arising from non-Hermiticity. According to Equation \eqref{equation6}, apart from a finite constant, the scattering loss $\mathrm{Im}\left(\Sigma_k\right)$ in a non-Hermitian quadratic band also has different exponents at the next leading order in $k$, as a function of the hyperuniformity exponent, $\alpha$. When $\alpha>2$, the scaling exponent $\beta_2$ of the scattering loss $\mathrm{Im}(\Sigma_\mathbf{k})=C_0+C_{\beta_2}\cdot k^{\beta_2}$ always satisfies $\beta_2=2$. Interestingly, there is a sudden jump in the next leading-order exponent at $\alpha=\frac{2}{\pi}\arctan\left(\frac{\mathrm{Re}(m)}{\mathrm{Im}(m)}\right)+1$. This is because the coefficient of the $k^\alpha$ term in Equation \eqref{equation6} accidentally goes to zero and the third leading order $O(k^2)$ becomes the second leading order. More details can be found in Supplemental Information Section 4. Figure \ref{figure4}(d) compares this second leading-order exponent $\beta_2$ between the theoretical analysis (black solid line, calculated from Equation \eqref{equation6}) and TB simulations (red dots, obtained by fitting the scattering loss $\mathrm{Im}\left(\Sigma_k\right)$ from $C_0+C_{\beta_2}\cdot k^{\beta_2}$, with three fitting parameters $C_0$, $C_{\beta_2}$ and $\beta_2$). Here, the FDTD simulation data are not shown in the figure, because the small fluctuations arising from finite-size effects destroy the second leading-order behavior as shown in Figure \ref{figure4}(b). Compared with the Hermitian case, which is shown in the figure by a dashed line, the second-order exponent in TB simulations is well captured by the non-Hermitian theory. Due to the higher-order effects arising from multiple scattering (as explained earlier), it is difficult to observe the sudden jump at $\alpha=\frac{2}{\pi}\arctan\left(\frac{\mathrm{Re}(m)}{\mathrm{Im}(m)}\right)+1$. For large $\alpha$, the TB simulation exceeds the upper bound of $\beta_2=2$ predicted in the theory. This is because $\beta_2$ is fitted within a finite range of $k_x\in [0, 0.04] \ [2\pi a^{-1}]$, so the higher-order exponent of $k$ enters the fitting process at $k\neq 0$, making $\beta_2$ slightly higher than theoretical predictions.

\section*{3\quad Conclusion}

In conclusion, we have investigated the wave dynamics of hyperuniform disorder in photonic crystal slabs, with a focus on non-Hermiticity that arises due to radiation loss out of the plane of the slab. By using excess linewidth as an observable to quantify the scattering loss introduced by disorder,  we characterize the wave dynamics of hyperuniform disorder by its strength of scattering at different wavevector $\mathbf{k}$. We theoretically derive the scattering loss associated with the imaginary part of the self-energy $\mathrm{Im}\left(\Sigma_k\right)$ in a Hermitian quadratic band and a non-Hermitian quadratic band with a complex effective mass, and verify the results in TB and FDTD simulations with realistic photonic crystal parameters. 

For hyperuniform disorder with hyperuniformity exponent $\alpha$, the strength of scattering in a Hermitian quadratic band is characterized by a power law behavior of $\mathrm{Im}(\Sigma_k)\propto k^\alpha$, which is observed and verified in a TB simulation. 

In contrast, the non-Hermiticity that arises naturally due to radiative loss significantly alters the scattering behavior of disorder. We show theoretically that the strength of scattering in a non-Hermitian quadratic band is always a finite constant (independent of k) at leading order. This finite constant is proportional to $\frac{\alpha+2}{\alpha}$, which decreases with $\alpha$, indicating the suppression of scattering due to hyperuniformity.  In addition, the exponent of the second leading term in $k$, labeled $k^{\beta_2}$, exhibits very different properties as a function of $\alpha$ compared to the Hermitian case.  For example, it never exceeds $\beta_2\leq 2$, no matter how large the hyperuniformity exponent $\alpha$ is, showing a clear difference from the Hermitian case.  The above results are observed and verified in both TB simulations and FDTD simulations. In the simulations, the scattering at small $\alpha$ is generally smaller than the theoretical predictions, which is attributed to higher-order effects arising from multiple scattering. We employ the SCBA method to consider multiple-scattering events and observe quantitative agreement between the SCBA numerical calculations and simulations.

The results presented here provide an understanding of the dynamics of hyperuniform disorder in photonic crystal slabs, making clear that non-Hermiticity must be incorporated for an accurate description of wave dynamics.  As such, these findings serve as a benchmark for the interplay between hyperuniform disorder and non-Hermiticity. The theoretical framework developed here can be widely applied to various disorder settings under different band dispersions, providing a starting point for exploring disorder effects in non-Hermitian systems more broadly. The FDTD simulations used in this work are designed with realistic photonic crystal parameters, leading to possible experimental realizations. Finally, the wave dynamics of hyperuniform disorder in non-Hermitian systems provides insights for practical device design, since non-Hermiticity arising from loss or dissipation is ubiquitous in realistic systems.

\section*{Acknowledgments}
We acknowledge the Army Research Office (ARO) under the MURI program, grant number W911NF-22-2-0103 and the Air Force Office of Scientific Research (AFOSR) under grant number FA9550-22-1-0339.

\clearpage
\onecolumngrid

\clearpage

\titleformat{\section}
  {\normalfont\normalsize\bfseries\centering}
  {\thesection}              
  {1em}                       
  {}   

\newgeometry{margin=2cm}
\section*{Supplemental Information for: \texorpdfstring{\\ Hyperuniform Disorder in Photonic Crystal Slabs with Intrinsic non-Hermiticity}{}}

\label{supplementary}

\renewcommand{\theequation}{S\arabic{equation}}
\renewcommand{\thefigure}{S\arabic{figure}}
\setcounter{equation}{0}
\setcounter{figure}{0}

\tableofcontents
\clearpage
\restoregeometry


\phantomsection
\addcontentsline{toc}{section}{Overview}

\section*{Overview} 

This Supplementary Information section acts as a companion to the main text, where we provide additional details on theoretical derivations.\\

In Section 1, we give a brief introduction to the general formalism of disorder. We present the definition of the spectral function $A_\mathbf{k}(E)$ and explain how it is broadened by the imaginary part of self-energy in a disordered system, which is related to the scattering loss. Next, we show the relation between the spectral function and the linewidth of a mode in a photonic crystal system.\\

In Section 2, we provide more details on how we use the Fourier filtering method to generate a hyperuniform disorder configuration in the context of photonic crystals. The local potential distribution forms a hyperuniform configuration. The hyperuniform pattern is characterized by a hyperuniformity exponent $\alpha$, such that the spectral density $\tilde{\rho}(\mathbf{q})\to q^\alpha$ as $q\to 0$ ($q=\left|\mathbf{q}\right|$ represents the magnitude of the wavevector).\\

In Section 3, we analytically calculate the scattering loss and the excess linewidth introduced by hyperuniform disorder in a Hermitian quadratic band. We find that the scattering loss is always in the form of $\mathrm{Im}(\Sigma_\mathbf{k})\to k^\alpha$, where $\alpha$ is the hyperuniformity exponent and $k=|\mathbf{k}|$.\\

In Section 4, we extend the theory to a non-Hermitian quadratic band with complex effective mass. We analytically calculate the scattering loss and the excess linewidth in this non-Hermitian setting. We find that when the hyperuniformity exponent $\alpha>0$, the leading order of the excess linewidth is always in the form of $C_0+C_\beta\cdot k^\beta$, where $C_0$ is a finite constant, and the exponent of the next leading order never exceeds $\beta\leq 2$. This clearly shows the fact that non-Hermiticity significantly alters the wave dynamics in hyperuniform systems.

\clearpage

\phantomsection
\addcontentsline{toc}{section}{Section 1: Formalism and notation: from disorder to photonic crystal}
\section*{Section 1: Formalism and notation: from disorder to photonic crystal}

In this section, we introduce the notation of disorder and the basic formalism for adding disorder in the context of photonic crystals.

The eigenmodes of a two-dimensional photonic crystal slab are governed by the Maxwell's equations, which lead to the following eigenvalue problem:
\begin{equation}
\mathcal{L} \mathbf{H_k}=E_\mathbf{k} \mathbf{H_k},
\label{a1}
\end{equation}
where $\mathcal{L}=\mathbf{\nabla}\times \left[\varepsilon(\mathbf{r})^{-1}\mathbf{\nabla}\times\cdot\right]$ is the Maxwell operator with $\varepsilon(\mathbf{r})$ the dielectric function, $\mathbf{r}=(x,y,z)$ is the three spactial coordinates, $\mathbf{k}=\left(k_x, k_y\right)$ is the 2D lattice Bloch momentum, $\mathbf{H_k}$ is the magnetic field profile of the eigenmode, and $E_\mathbf{k}=\left(\frac{\omega_\mathbf{k}}{c}\right)^2$ is the energy of the eigenmode, where $\omega_\mathbf{k}$ is the eigenfrequency of the mode, and $c$ is the speed of light. 

Two-dimensional photonic crystal slabs support both guided modes confined below the light line and leaky resonances that can couple to the radiation continuum \cite{fan2002analysis,photonicbook}. In this work, we focus on radiative modes that are above the light line, centered at $\mathbf{k}=\Gamma=(0,0)$. Because of their radiative coupling, these modes are characterized by finite lifetimes, which can be incorporated into the formalism through a complex eigenvalue $E_\mathbf{k}$. Throughout this paper, the imaginary part of the eigenvalue $\mathrm{Im}\left(E_\mathbf{k}\right)$ in the periodic lattice is referred to as the intrinsic loss (or intrinsic linewidth), in order to distinguish it from the additional broadening induced by disorder scattering (or excess linewidth). For notation simplicity and analytical convenience, we describe band dispersions in terms of the eigenvalue 
$E=\left(\omega/c\right)^2$ of the Maxwell's equations rather than the frequency $\omega$ itself. By analogy with the dispersion of a massive particle in quantum mechanics, this quantity is referred to as ``energy'', despite it has a unit of $[a^{-2}]$.

The Green's function in the periodic lattice can be defined as:
\begin{equation}
G^0_{\mathbf{k},\mathbf{k'}}(E)=\left<\mathbf{H_k}\left|\frac{1}{E-\mathcal{L}+i0^{+}}\right|\mathbf{H_{k'}}\right>.
\label{a2}
\end{equation}

For simplicity of notation, the $i0^{+}$ term is omitted from this point onward. The spectral function in the periodic lattice is defined as \cite{fetter2012quantum}:
\begin{equation}
A_{\mathbf{k}}^{0}(E)=\mathrm{Im}\left(G^0_{\mathbf{k},\mathbf{k}}(E)\right).
\label{a3}
\end{equation}

From Eq. \eqref{a1}, we can see that the spectral function can be written in terms of the real and imaginary parts of the energy $E_\mathbf{k}$:
\begin{equation}
A_{\mathbf{k}}^{0}(E)=\mathrm{Im}\left(\frac{1}{E-E_\mathbf{k}}\right)=\frac{\mathrm{Im}\left(E_\mathbf{k}\right)}{\left(E-\mathrm{Re}\left(E_\mathbf{k}\right)\right)^2+\left(\mathrm{Im}\left(E_\mathbf{k}\right)\right)^2},
\label{a4}
\end{equation}
which means $A^{0}_{\mathbf{k}}(E)$ forms a Lorentzian lineshape where the peak is centered at $E=\mathrm{Re}\left(E_\mathbf{k}\right)$, with full width at half maximum (FWHM) $-2\mathrm{Im}\left(E_\mathbf{k}\right)$.

Then, a weak disorder $V(\mathbf{r})$ that breaks the translational symmetry is added to the system; in other words, $\mathbf{k}$ is no longer a good quantum number. The spectral density of the system can be defined as:
\begin{equation}
\tilde{\rho}(\mathbf{q})=\left<\tilde{V}(\mathbf{q})\tilde{V}(\mathbf{-q})\right>_{\mathrm{dis}},
\label{a5}
\end{equation}
where $\tilde{V}(\mathbf{q})$ is the Fourier component of $V(\mathbf{r})$ corresponding to momentum $\mathbf{q}$, and $\left<\cdot\right>_{\mathrm{dis}}$ represents the average over disorder configurations.

According to disorder theory \cite{lee1985disordered}, the averaged Green’s function can always be written in terms of self-energy:
\begin{equation}
\left<G_{\mathbf{k},\mathbf{k}}\left(E\right)\right>_{\mathrm{dis}}=\frac{1}{\left(E-E_\mathbf{k}\right)-\Sigma_\mathbf{k}(E)}, 
\label{a6}
\end{equation}
where $\Sigma_\mathbf{k}(E)$ is the self-energy.

In the weak disorder limit, the self-energy can be calculated perturbatively:
\begin{equation}
\Sigma_\mathbf{k}(E)=\frac{1}{N_x N_y}\sum_\mathbf{q} G^0_{\mathbf{k+q},\mathbf{k+q}}(E) \tilde{\rho}(\mathbf{q})+O(\tilde{\rho}^2),
\label{a7}
\end{equation}
where $N_x$ and $N_y$ are the numbers of unit cells in the $x$ and $y$ directions. The first term is the leading-order contribution proportional to the spectral density $\tilde{\rho}(\mathbf{q})$, which represents a single-scattering event from $\mathbf{k}$ to $\mathbf{k+q}$. The next leading term is $O(\tilde{\rho}^2)$, which represents double-scattering events, and so on.

From Eq. \eqref{a6}, the spectral function in the disorder system becomes:
\begin{equation}
A_{\mathbf{k}}(E)
=\mathrm{Im}\left<G_{\mathbf{k},\mathbf{k}}\left(E\right)\right>_{\mathrm{dis}}
=\frac{\mathrm{Im}\left(E_\mathbf{k}\right)+\mathrm{Im}\left(\Sigma_\mathbf{k}(E)\right)}
{\left(E-\mathrm{Re}\left(E_\mathbf{k}\right)-\mathrm{Re}\left(\Sigma_\mathbf{k}(E)\right)\right)^2+\left(\mathrm{Im}\left(E_\mathbf{k}\right)+\mathrm{Im}\left(\Sigma_\mathbf{k}(E)\right)\right)^2},
\label{a8}
\end{equation}
After adding disorder, the spectral function still has a Lorentzian lineshape, but the center energy is shifted to $E=\mathrm{Re}\left(E_\mathbf{k}\right)+\mathrm{Re}\left(\Sigma_\mathbf{k}(E)\right)$, with FWHM $-2\mathrm{Im}\left(E_\mathbf{k}\right)-2\mathrm{Im}\left(\Sigma_\mathbf{k}(E)\right)$. Compared with Eq. \eqref{a4}, we can see that the width of the peak is broadened from $-2\mathrm{Im}\left(E_\mathbf{k}\right)$ in the non-disordered case (which is referred to as intrinsic linewidth) to $-2\mathrm{Im}\left(E_\mathbf{k}\right)-2\mathrm{Im}\left(\Sigma_\mathbf{k}(E)\right)$ after adding disorder. We refer to the extra linewidth, $-2\mathrm{Im}\left(\Sigma_\mathbf{k}(E)\right)$, as the linewidth broadening or excess linewidth, which is directly caused by the scattering loss from disorder.

As we can see, after a disorder is added to the system, the translational symmetry is broken by the disorder, so the wavevectors $\mathbf{k}$ are no longer conserved. The band structure is not well-defined. However, the spectral property of the system can still be described by the spectral function defined in Eq. \eqref{a8}.

The energy $E_\mathbf{k}$ and the frequency $\omega_\mathbf{k}$ can be written in their real part and imaginary part:
$E_\mathbf{k}=\mathrm{Re}(E_\mathbf{k})+i\cdot\mathrm{Im}(E_\mathbf{k})$ and $\omega_\mathbf{k}=\mathrm{Re}(\omega_\mathbf{k})+i\cdot\mathrm{Im}(\omega_\mathbf{k})$. Given that $E_\mathbf{k}=\left(\frac{\omega_\mathbf{k}}{c}\right)^2$ and assuming that $\mathrm{Re}(\omega_\mathbf{k})\gg \mathrm{Im}(\omega_\mathbf{k})$, we get:
\begin{equation}
\mathrm{Re}(E_\mathbf{k})=\frac{\mathrm{Re}(\omega_\mathbf{k})^2}{c^2},
\label{a9}
\end{equation}
\begin{equation}
\mathrm{Im}(E_\mathbf{k})=\frac{2\cdot\mathrm{Re}(\omega_\mathbf{k})\cdot\mathrm{Im}(\omega_\mathbf{k})}{c^2}.
\label{a10}
\end{equation}

In order to find the center wavelength ($\lambda_0$) and linewidth ($\Delta \lambda$) of the mode, we notice that the ratio between the linewidth and the resonant value remains the same in wavelength and frequency:
\begin{equation}
\frac{\Delta \lambda}{\lambda_0}=\left|\frac{\Delta \omega}{\omega_0}\right|=2\left|\frac{\mathrm{Im}(\omega)}{\mathrm{Re}(\omega)}\right|,
\label{a11}
\end{equation}
\begin{equation}
\lambda_0=\frac{2\pi c}{\omega_0}.
\label{a12}
\end{equation}

Writing the linewidth of wavelength $\Delta \lambda$ in terms of energy, we find: 
\begin{equation}
\Delta \lambda=-\frac{2\pi \cdot \mathrm{Im}(E)}{\mathrm{Re}(E)^{\frac{3}{2}}}.
\label{a13}
\end{equation}

In the limit of weak disorder, the real part of the intrinsic energy is much larger than the real part of the self-energy; that is, $\mathrm{Re}\left(E_\mathbf{k}\right)\gg \mathrm{Re}\left(\Sigma_\mathbf{k}(E)\right)\propto \left\|V\right\|^2$, where $\left\|V\right\|$ represents the degree of disorder. In this limit, the scattering loss $\mathrm{Im}(\Sigma_\mathbf{k}(E))$ can be evaluated at the band structure in the periodic lattice where $E=\mathrm{Re}(E_\mathbf{k})$:
\begin{equation}
\Sigma_\mathbf{k}=\left.\Sigma_\mathbf{k}(E)\right|_{E=\mathrm{Re}(E_\mathbf{k})}.
\label{a14-0}
\end{equation}

According to Eqs. \eqref{a2} and \eqref{a7}:
\begin{equation}
\Sigma_\mathbf{k}=\frac{1}{N_x N_y}\sum_\mathbf{q} \frac{\tilde{\rho}(\mathbf{q})}{\mathrm{Re}\left(E_{\mathbf{k}}\right)-E_{\mathbf{k+q}}+i0^{+}}+O(\tilde{\rho}^2).
\label{a14-1}
\end{equation}

Equation \eqref{a14-1} is the same as Equation (4) in the main text.

According to Eq. \eqref{a13}, the excess linewidth (the difference between total linewidth and intrinsic linewidth) is:
\begin{equation}
\Delta \lambda_\mathbf{k}
=
-\frac{2\pi\cdot\mathrm{Im} \left(\Sigma_\mathbf{k}\right)}{\mathrm{Re}(E_\mathbf{k})^{\frac{3}{2}}}.
\label{a14}
\end{equation}

Next, we will use Eq. \eqref{a7} and Eq. \eqref{a14} to calculate the scattering loss $\mathrm{Im}(\Sigma_\mathbf{k})$ and the excess linewidth $\Delta\lambda_\mathbf{k}$, for hyperuniform disorder configurations in both Hermitian quadratic bands and non-Hermitian quadratic bands.

\clearpage

\phantomsection
\addcontentsline{toc}{section}{Section 2: Fourier filtering and spectral density}
\section*{Section 2: Fourier filtering and spectral density}

In this section, we introduce how the hyperuniform disorder configuration is generated by the Fourier filtering method and calculate the spectral density $\tilde{\rho}(\mathbf{q})$ as a function of disorder.

\begin{figure}[H]
    \centering
    \subfigure[]{\includegraphics[width=0.23\textwidth]{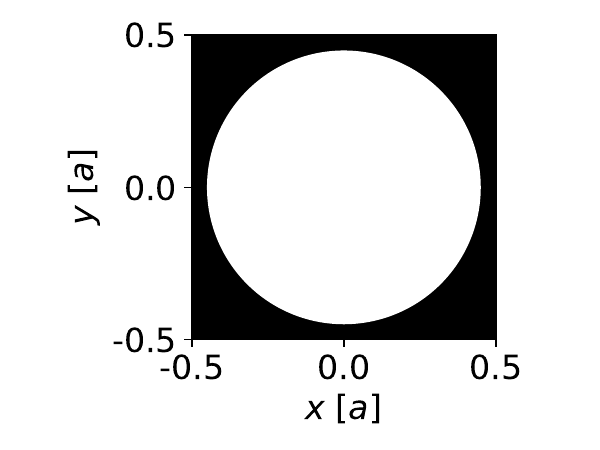}}
    \subfigure[]{\includegraphics[width=0.23\textwidth]{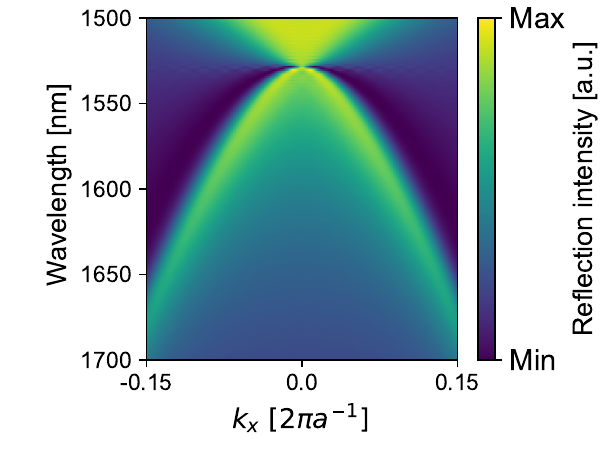}}
    \subfigure[]{\includegraphics[width=0.23\textwidth]{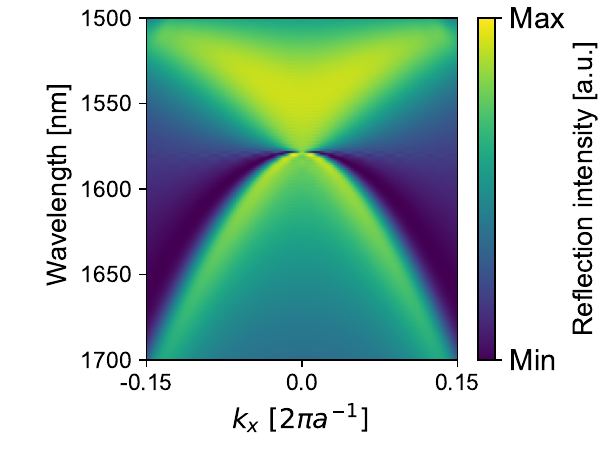}}
    \subfigure[]{\includegraphics[width=0.23\textwidth]{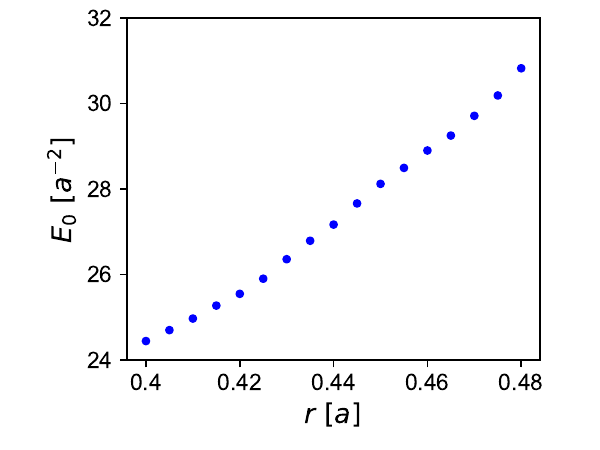}}

\caption{The method to add disorder into a photonic crystal slab. (a) The unit cell (in the $x-y$ plane, $z$ is the out-of-plane direction) of our photonic crystal slab. The structure contains a circular air hole ($\varepsilon=1.0$) with $r_0=0.45a$ in a square lattice, where $a=1290nm$ is the lattice constant. The slab is made of silicon ($\varepsilon=12.11$) with thickness $h=0.05a$. (b) The simulated band structure of (a) along the $k_y=0$ line. (c) The simulated band structure of (a), but with $r=0.43a$. (d) The tip energy ($E=\left(\frac{\omega}{c}\right)^2$) of the quadratic band as a function of the air hole radius $r$.}
\label{figs1}
\end{figure}

Figure \ref{figs1}(a) shows the photonic crystal slab structure used in this work. The structure contains circular air holes arranged to form a square lattice in a silicon slab. The structure hosts an isolated quadratic band in the vicinity of $\mathbf{k}=\Gamma$. To probe the band in the simulation, we illuminate the sample with a focused Gaussian beam, and collect the reflection intensity at the far field with different angles. Similar to the experimental setting (back focal plane imaging setup) in Ref. \cite{barsukova2024direct}, we obtain the spectral information of all the wavevector $\mathbf{k}$ within a single simulation. The simulation results are obtained from \textsc{Tidy3D}, a package developed by \textsc{Flexcompute} that uses the Finite-Difference Time-Domain (FDTD) method \cite{Minkov:24}.

In the simulation, we find that the band tip energy changes as we vary the radius of the air hole. In Fig. \ref{figs1}(b), the radius of the hole is $r=0.45a$, which is the radius in the periodic structure shown in Fig. 1 in the main text. The band tip energy is $E_0=28.1a^{-2}$ (corresponding to wavelength $\lambda_0=1529nm$). While in Fig. \ref{figs1}(c), the radius of the hole reduces to $r=0.43a$. The band tip energy decreases to $E_0=26.4a^{-2}$ (corresponding to wavelength $\lambda_0=1578nm$). More generally, we show in Fig. \ref{figs1}(d) that the tip energy increases as the radius of the hole increases. 

Figure \ref{figs1} gives us the intuition to introduce disorder by varying the size of the hole at each site. A potential disorder can be added to a photonic crystal slab by varying the radii of the holes $r_{l,j}$ at each site $(l,j)$, such that the local potential (which is defined as the tip energy corresponding to $r_{l,j}$ in the periodic structure) $V_{l,j}$ forms a disorder pattern. This approach allows us to quantitatively assign a value to the local potential $V_{l,j}$ that is directly connected to the radius of the hole $r_{l,j}$ at each site. By carefully designing the radius of the hole $r_{l,j}$ at each site, we are able to add a controllable potential disorder $V_{l,j}$ to a photonic crystal slab.

To generate a hyperuniform disorder configuration, we first start from a random, uncorrelated disorder pattern, such that the local potential $V_{l,j}$ is randomly and uniformly selected from the range 
\begin{equation}
V_{l,j} \sim U\left(E_0-w, E_0+w\right),
\label{d1}
\end{equation}
where $U$ represents a uniform distribution, $E_0=28.1a^{-2}$ is the tip energy in the periodic lattice, and $w$ denotes the degree of disorder.

Then, the Fourier component of the local potential $V_{l,j}$ is calculated by discrete Fourier transform:
\begin{equation}
\tilde{V}(\mathbf{q})=\frac{1}{\sqrt{N_x N_y}}\sum_{l,j}e^{-i \left(l q_x a+j q_y a\right)} V_{l,j},
\label{d2}
\end{equation}
where $N_x$ and $N_y$ are the number of sites in the $x$ and $y$ directions, and $\mathbf{q}=\left(q_x,q_y\right)$ represents the Fourier component.

Note that $V_{l,j}\in\mathbb{R}$, which means $\tilde{V}(-\mathbf{q})=\left(\tilde{V}(\mathbf{q})\right)^{*}$, or equivalently, $\tilde{V}(-\mathbf{q})\tilde{V}(\mathbf{q})=\left|\tilde{V}(\mathbf{q})\right|^2$. 

It is straightforward to prove that both the real part and the imaginary part of $\tilde{V}\left(\mathbf{q}\right)$ satisfy a normal distribution for large system size $N_x\times N_y$:
\begin{equation}
\mathrm{Re} \left(\tilde{V}(\mathbf{q})\right)\sim N\left(0,\frac{1}{6}w^2\right),
\label{d3}
\end{equation}
\begin{equation}
\mathrm{Im} \left(\tilde{V}(\mathbf{q})\right)\sim N\left(0,\frac{1}{6}w^2\right),
\label{d4}
\end{equation}
where $N$ denotes a normal distribution. Therefore,
\begin{equation}
\tilde{\rho}(\mathbf{q})
=\left<\tilde{V}(-\mathbf{q}) \tilde{V}(\mathbf{q)}\right>=
\left<\left|\tilde{V}(\mathbf{q})\right|^2\right>=\frac{1}{3}w^2.
\label{d5}
\end{equation}

Next, we employ a Fourier filtering method \cite{makse1996method}, such that all the Fourier components $\tilde{V}(\mathbf{q})$ are modified and filtered in the following way:
\begin{equation}
\tilde{V}^{\prime}(\mathbf{q})=
\left\{
\begin{aligned}
\sqrt{\frac{\alpha+2}{2}}\left(\frac{q}{q_\mathrm{max}}\right)^{\frac{\alpha}{2}}\tilde{V}(\mathbf{q}) & , & q<q_\mathrm{max}\\
0 & , & q>q_\mathrm{max}
\end{aligned}
\right.
\label{d6}
\end{equation}

Here, $\tilde{V}^{\prime}(\mathbf{q})$ denotes the Fourier component after the Fourier filtering, $q_\mathrm{max}=0.3\ [2\pi a^{-1}]$ is a cutoff wavenumber to make sure that the Fourier components are isotropic in the first Brillouin zone, and $\alpha$ is the hyperuniformity exponent that we will show later. After the filtering, the spectral density $\tilde{\rho}(\mathbf{q})$ becomes:
\begin{equation}
\tilde{\rho}(\mathbf{q})=
\left\{
\begin{aligned}
\frac{\alpha+2}{6}w^2 \left(\frac{q}{q_\mathrm{max}}\right)^\alpha & , & q<q_\mathrm{max}\\
0 & , & q>q_\mathrm{max}
\end{aligned}
\right.
\label{d7}
\end{equation}

The spectral density shown in Eq. \eqref{d7} represents a hyperuniform disorder with a hyperuniformity exponent $\alpha$, because $\tilde{\rho}(\mathbf{q})\to q^\alpha$ when $q\to 0$. Due to the properties of the discrete Fourier transform, $\tilde{\rho}(\mathbf{q})$ should be periodic in both $q_x$ and $q_y$ directions: $\tilde{\rho}\left(q_x+\frac{2\pi}{a},q_y\right)=\tilde{\rho}\left(q_x,q_y+\frac{2\pi}{a}\right)=\tilde{\rho}\left(q_x,q_y\right)$. Equation \eqref{d7} is the same as Eq. (3) in the main text.

Finally, the inverse discrete Fourier transformation is performed to transform $\tilde{V}^{\prime}(\mathbf{q})$ back into real space:
\begin{equation}
V^{\prime}_{l,j}=\frac{1}{\sqrt{N_x N_y}}\sum_{\mathbf{q}}e^{i \left(l q_x a + j q_y a\right)} \tilde{V}^{\prime}(\mathbf{q}).
\label{d8}
\end{equation}

Since the filtering in Eq. \eqref{d6} does not change the fact that $\tilde{V}^{\prime}(-\mathbf{q})=\left(\tilde{V}^{\prime}(\mathbf{q})\right)^{*}$, the local potential after the filtering is still a real number $V^{\prime}_{l,j}\in\mathbb{R}$.

According to Eqs. \eqref{d5}, \eqref{d6}, and \eqref{d8}, the variance of the local potential after the Fourier filtering is:
\begin{equation}
\mathrm{Var}\left(V_{l,j}^\prime\right)
=
\frac{1}{3}w^2\left(\frac{a}{2\pi}\right)^2
\cdot
\iint_{q<q_\mathrm{max}} d\mathbf{q}\cdot\frac{\alpha+2}{2}\left(\frac{q}{q_\mathrm{max}}\right)^{\alpha}
=\frac{a^2 q_\mathrm{max}^2w^2}{12\pi}
\label{d9}
\end{equation}

We can see that the variance of the local potential in real space $\mathrm{Var}\left(V^\prime_{l,j}\right)$ is a constant independent of $\alpha$. This allows us to make fair comparisons across the results from different hyperuniformity exponents $\alpha$.

\clearpage

\phantomsection
\addcontentsline{toc}{section}{Section 3: Scattering loss and excess linewidth in a Hermitian quadratic band}
\section*{Section 3: Scattering loss and excess linewidth in a Hermitian quadratic band}

In this section, we analytically calculate the scattering loss $\mathrm{Im}(\Sigma_\mathbf{k})$ and the excess linewidth $\Delta\lambda_\mathbf{k}$ introduced by disorder when we have a Hermitian quadratic band in a lattice system.

For a Hermitian quadratic band in a lattice, the band dispersion is 
\begin{equation}
E_\mathbf{k}=E_0-\frac{1}{2m}\left(k_x^2+k_y^2\right)\quad 
\left(-\frac{\pi}{a}<k_x, k_y<\frac{\pi}{a}\right)
\label{b1}
\end{equation}
where $a$ is the lattice constant, $E_0\in \mathbb{R}$ is the tip energy of the quadratic band, $m\in\mathbb{R}$ is the effective mass, and the band structure is periodic under $k_x\mapsto k_x+\frac{2\pi}{a}$ and $k_y\mapsto k_y+\frac{2\pi}{a}$. For the photonic band used in this work, $\mathrm{Re}(m)>0$, so we assume $m>0$ in this section.

Next, we consider a 2D photonic crystal slab of circular holes in a square lattice. By changing the radius of the hole at site $\left(l,j\right)$, the local potential $V_{l,j}$ also changes. We use the Fourier filtering method to add a hyperuniform disorder to the system, as explained in detail in Section 2, such that the spectral density satisfies:
\begin{equation}
\tilde{\rho}(\mathbf{q})=
\left\{
\begin{aligned}
\frac{\alpha+2}{6}w^2 \left(\frac{q}{q_\mathrm{max}}\right)^\alpha & , & q<q_\mathrm{max}\\
0 & , & q>q_\mathrm{max}
\end{aligned}
\right.
\label{b2}
\end{equation}
where $\alpha\geq 0$ is the hyperuniformity exponent, $w$ is the disorder strength, and $q_\mathrm{max}=0.3\ [2\pi a^{-1}]$ is a cutoff wavenumber to make the spectral density symmetric in the first Brillouin zone.

Notice that:
\begin{equation}
\frac{1}{N_x N_y}\sum_\mathbf{q}=\left(\frac{a}{2\pi}\right)^2\iint d\mathbf{q}
\label{b4}
\end{equation}

From Eqs. \eqref{a7}, \eqref{a14-0}, \eqref{b2}, and \eqref{b4}, we get:
\begin{equation}
\begin{aligned}
\Sigma_\mathbf{k}
&=\frac{1}{N_x N_y}\sum_\mathbf{q}\frac{\tilde{\rho}(\mathbf{q})}{\mathrm{Re}\left(E_\mathbf{k}\right)-E_{\mathbf{k+q}}}\\
&=\frac{1}{3} w^2 \frac{\alpha+2}{2} \left(\frac{1}{q_\mathrm{max}}\right)^\alpha \left(\frac{a}{2\pi}\right)^2\iint_{q<q_\mathrm{max}}d\mathbf{q}\frac{q^\alpha}{-\frac{k^2}{2m}+\frac{\left(\mathbf{k+q}\right)^2}{2m}}\\
&=\frac{1}{3} w^2 \frac{\alpha+2}{2} \left(\frac{1}{q_\mathrm{max}}\right)^\alpha\left(\frac{a}{2\pi}\right)^2\frac{m}{k}\int_0^{q_\mathrm{max}}dq\ q^\alpha\int_0^{2\pi}d\theta\frac{1}{-\frac{q}{2k}-\cos\theta}.
\end{aligned}
\label{b5}
\end{equation}

Notice the fact that:
\begin{equation}
\int_0^{2\pi}\frac{1}{x-cos\theta}=
\left\{
\begin{aligned}
\frac{2\pi}{\sqrt{x^2-1}} & , & x>1\\
-\frac{2\pi}{\sqrt{x^2-1}} & , & x<-1\\
i\frac{2\pi}{\sqrt{1-x^2}} & , & |x|<1\\
\end{aligned}
\right.
\label{b6}
\end{equation}

Equation \eqref{b5} becomes:
\begin{equation}
\begin{aligned}
\Sigma_\mathbf{k}
&=-\frac{1}{3} w^2 \frac{\alpha+2}{2} \left(\frac{1}{q_\mathrm{max}}\right)^\alpha\left(\frac{a}{2\pi}\right)^2\frac{m}{k}
\left(
\int_0^{2k}2\pi i\frac{q^\alpha}{\sqrt{1-\left(\frac{q}{2k}\right)^2}}dq
-
\int_{2k}^{q_\mathrm{max}}2\pi \frac{q^\alpha}{\sqrt{\left(\frac{q}{2k}\right)^2-1}}dq
\right)\\
&=
-\frac{1}{3} w^2 \frac{\alpha+2}{2} \left(\frac{2k}{q_\mathrm{max}}\right)^\alpha\left(\frac{a}{2\pi}\right)^2 2m\cdot 2\pi
\left(
i\int_0^{1} \frac{x^\alpha}{\sqrt{1-x^2}}dx
-
\int_{1}^{\frac{q_\mathrm{max}}{2k}} \frac{x^\alpha}{\sqrt{x^2-1}}dx
\right)
.
\end{aligned}
\label{b7}
\end{equation}

Here we already assume $k<\frac{q_\mathrm{max}}{2}$, because we are only interested in the behavior near $k=0$.

Since we only care about the imaginary part of the self energy, and $m\in\mathbb{R}$, we only need to consider the contribution from the first integral. And also notice the fact that:
\begin{equation}
\int_0^1\frac{x^\alpha}{\sqrt{1-x^2}}dx=\frac{\sqrt{\pi}}{2}\frac{\Gamma\left(\frac{\alpha+1}{2}\right)}{\Gamma\left(\frac{\alpha+2}{2}\right)}.
\label{b8}
\end{equation}

We get:
\begin{equation}
\begin{aligned}
\mathrm{Im}\left(\Sigma_\mathbf{k}\right)
&=
-\frac{1}{3} w^2 \frac{\alpha+2}{2} \left(\frac{2k}{q_\mathrm{max}}\right)^\alpha\left(\frac{a}{2\pi}\right)^2 2m\cdot 2\pi
\int_0^{1} \frac{x^\alpha}{\sqrt{1-x^2}}dx
\\
&=
-\frac{2^\alpha(\alpha+2)\cdot\Gamma\left(\frac{\alpha+1}{2}\right)\cdot w^2 a^2 m}{12\sqrt{\pi}\cdot\Gamma\left(\frac{\alpha+2}{2}\right)\cdot q_\mathrm{max}^\alpha}k^\alpha.
\end{aligned}
\label{b9}
\end{equation}

Equation \eqref{b9} is the same as Eq. (5) in the main text.

According to Eq. \eqref{a14}:
\begin{equation}
\Delta\lambda_\mathbf{k}
=
\frac{\sqrt{\pi}\cdot2^\alpha(\alpha+2)\cdot\Gamma\left(\frac{\alpha+1}{2}\right)\cdot w^2 a^2 m}{6\cdot\Gamma\left(\frac{\alpha+2}{2}\right)\cdot q_\mathrm{max}^\alpha\cdot \mathrm{Re}\left(E_\mathbf{k}\right)^{\frac{3}{2}}}k^\alpha.
\label{b10}
\end{equation}

From Eqs. \eqref{b10} and \eqref{b10}, we can see that:

1. The scattering loss $\mathrm{Im}(\Sigma_\mathbf{k})$ and the excess linewidth $\Delta\lambda_\mathbf{k}$ are proportional to $k^\alpha$

2. When $\alpha=0$, the scattering loss and excess linewidth both reduce to the uncorrelated disorder, such that they are constants independent of $k$:
\begin{equation}
\mathrm{Im}(\Sigma_\mathbf{k})
=
-\frac{1}{6}w^2 a^2 m,
\label{b11-0}
\end{equation}
\begin{equation}
\Delta\lambda_\mathbf{k}
=
\frac{\pi w^2 a^2 m}{3\cdot \mathrm{Re}(E_\mathbf{k})^{\frac{3}{2}}}.
\label{b11}
\end{equation}

In Eq. \eqref{b11}, we notice that the scattering loss $\mathrm{Im}(\Sigma_\mathbf{k})$ and the excess linewidth $\Delta\lambda_\mathbf{k}$ are independent of the cutoff wavenumber $q_\mathrm{max}$. This is because in the Hermitian case, the energy is conserved. The single scattering is only allowed from one point on the iso-frequency contour to another point on the IFC. Hence, $q_\mathrm{max}$ has no influence when $k<\frac{q_{\mathrm{max}}}{2}$.

\clearpage

\phantomsection
\addcontentsline{toc}{section}{Section 4: Scattering loss and excess linewidth in a non-Hermitian quadratic band with complex effective mass}
\section*{Section 4: Scattering loss and excess linewidth in a non-Hermitian quadratic band with complex effective mass}

In this section, we will extend our analytical derivation of the scattering loss $\mathrm{Im}(\Sigma_\mathbf{k})$ and the excess linewidth $\Delta\lambda_\mathbf{k}$ to a non-Hermitian quadratic band with complex effective mass.

In the previous section, we consider a Hermitian quadratic band:
\begin{equation}
E_\mathbf{k}=E_0-\frac{1}{2m}\left(k_x^2+k_y^2\right)\quad 
\left(-\frac{\pi}{a}<k_x, k_y<\frac{\pi}{a}\right)
\label{c1}
\end{equation}
where the effective mass is real $m\in \mathbb{R}$. In this section, we will consider the case that $m$ is no longer a pure real number, but with a non-zero imaginary part:
\begin{equation}
m=\mathrm{Re}(m)+i\cdot \mathrm{Im}(m)
\label{c2}
\end{equation}

From this point onward, we will only consider the case $\mathrm{Re}(m)>0$ and $\mathrm{Im}(m)<0$, which is true in the photonic crystal slab structure we are interested.

The derivation from Eq. \eqref{b1} to \eqref{b4} remains the same, but starting from Eq. \eqref{b5}, the imaginary part of the effective mass will play a role. The Eq. \eqref{b5} needs to be modified as:
\begin{equation}
\begin{aligned}
\Sigma_\mathbf{k}
&=\frac{1}{N_x N_y}\sum_\mathbf{q}\frac{\tilde{\rho}(\mathbf{q})}{\mathrm{Re}\left(E_\mathbf{k}\right)-E_{\mathbf{k+q}}}\\
&=\frac{1}{3} w^2 \frac{\alpha+2}{2} \left(\frac{1}{q_\mathrm{max}}\right)^\alpha \left(\frac{a}{2\pi}\right)^2\iint_{q<q_\mathrm{max}}d\mathbf{q}\frac{q^\alpha}{-\mathrm{Re}\left(\frac{k^2}{2m}\right)+\frac{\left(\mathbf{k+q}\right)^2}{2m}}\\
&=\frac{1}{3} w^2 \frac{\alpha+2}{2} \left(\frac{1}{q_\mathrm{max}}\right)^\alpha \left(\frac{a}{2\pi}\right)^2\iint_{q<q_\mathrm{max}}d\mathbf{q}\frac{q^\alpha}{-\frac{k^2}{2m}+\mathrm{Im}\left(\frac{k^2}{2m}\right)+\frac{\left(\mathbf{k+q}\right)^2}{2m}}.
\end{aligned}
\label{c3}
\end{equation}

We can see that there is an extra term $\mathrm{Im}\left(\frac{k^2}{2m}\right)$ in the denominator. Since we only care about the behavior near $k=0$, the approximation $\mathrm{Im}\left(\frac{k^2}{2m}\right)\approx0$ is made in the following derivation. This is equivalent to:
\begin{equation}
\mathrm{Re}\left(\frac{k^2}{2m}\right)\approx\frac{k^2}{2m}
\label{c4}    
\end{equation}

Under the above approximation, the derivations from Eq. \eqref{b5} to \eqref{b8} are still valid in our non-Hermitian case:
\begin{equation}
\begin{aligned}
\Sigma_\mathbf{k}
=
-\frac{1}{3} w^2 \frac{\alpha+2}{2} \left(\frac{2k}{q_\mathrm{max}}\right)^\alpha\left(\frac{a}{2\pi}\right)^2 \cdot2m\cdot 2\pi
\left(
\int_0^{1} \frac{x^\alpha}{\sqrt{1-x^2}}dx
-
\int_{1}^{\frac{q_\mathrm{max}}{2k}} \frac{x^\alpha}{\sqrt{x^2-1}}dx
\right)
\end{aligned}
\label{c5}
\end{equation}

However, since the effective mass $m$ is no longer a pure real number, we also need to consider the contribution from the second integral in Eq. \eqref{c5}. Hence,
\begin{equation}
\begin{aligned}
\mathrm{Im}\left(\Sigma_\mathbf{k}\right)
=
-\frac{1}{3} w^2 \frac{\alpha+2}{2} \left(\frac{2k}{q_\mathrm{max}}\right)^\alpha\left(\frac{a}{2\pi}\right)^2 \cdot2\cdot 2\pi
\left(
\mathrm{Re}(m)\cdot\int_0^{1} \frac{x^\alpha}{\sqrt{1-x^2}}dx
-
\mathrm{Im}(m)\cdot\int_{1}^{\frac{q_\mathrm{max}}{2k}} \frac{x^\alpha}{\sqrt{x^2-1}}dx
\right)
\end{aligned}
\label{c6}
\end{equation}

Notice the fact that:
\begin{equation}
\begin{aligned}
\int_1^y\frac{x^\alpha}{\sqrt{x^2-1}}dx
&=
\frac{\sqrt{\pi}}{2}\frac{\Gamma\left(-\frac{\alpha}{2}\right)}{\Gamma\left(\frac{1-\alpha}{2}\right)}
+\sum_{n=0}^\infty (-1)^n \binom{-\frac{1}{2}}{n}\frac{1}{\alpha-2n}\cdot y^{\alpha-2n}\\
&=
\frac{\sqrt{\pi}}{2}\frac{\Gamma\left(-\frac{\alpha}{2}\right)}{\Gamma\left(\frac{1-\alpha}{2}\right)}
+\frac{1}{\alpha}y^\alpha
+\frac{1}{2(\alpha-2)}y^{\alpha-2}
+O\left(y^{\alpha-4}\right)
\end{aligned}
\label{c7}
\end{equation}

Using the information in Eqs. \eqref{b8} and \eqref{c7}, Eq. \eqref{c6} becomes:
\begin{equation}
\scalebox{0.93}{$\displaystyle
\begin{aligned}
\mathrm{Im}\left(\Sigma_\mathbf{k}\right)&=
-\frac{(\alpha+2)w^2 V_0^2a^2}{6\pi}\left(\frac{2k}{q_\mathrm{max}}\right)^\alpha\times\\
&\quad\left(
\mathrm{Re}(m)\cdot\frac{\sqrt{\pi}}{2}\frac{\Gamma\left(\frac{\alpha+1}{2}\right)}{\Gamma\left(\frac{\alpha+2}{2}\right)}
-
\mathrm{Im}(m)\cdot\frac{\sqrt{\pi}}{2}\frac{\Gamma\left(-\frac{\alpha}{2}\right)}{\Gamma\left(\frac{1-\alpha}{2}\right)}
-
\frac{\mathrm{Im}(m)}{\alpha}\left(\frac{q_\mathrm{max}}{2k}\right)^\alpha
-\frac{\mathrm{Im}(m)}{2(\alpha-2)}\left(\frac{q_\mathrm{max}}{2k}\right)^{\alpha-2}
+O\left(\frac{q_\mathrm{max}}{2k}\right)^{\alpha-4}
\right)
\end{aligned}
\label{c8}
$}
\end{equation}

By carefully estimating the order (in $k$) of each term, we get:
\begin{equation}
\scalebox{0.86}{$\displaystyle
\begin{aligned}
\mathrm{Im}\left(\Sigma_\mathbf{k}\right)=&\frac{(\alpha+2)w^2a^2}{6\pi}\times\\
&\left\{
\begin{aligned}
-\mathrm{Im}(m)\cdot\ln\left(\frac{2k}{q_\mathrm{max}}\right)+O\left(k^0\right) & , & \alpha=0\\
\frac{\mathrm{Im}(m)}{\alpha}
+\left(
\mathrm{Im}(m)\cdot\frac{\sqrt{\pi}}{2}\frac{\Gamma\left(-\frac{\alpha}{2}\right)}{\Gamma\left(\frac{1-\alpha}{2}\right)}
-\mathrm{Re}(m)\cdot\frac{\sqrt{\pi}}{2}\frac{\Gamma\left(\frac{\alpha+1}{2}\right)}{\Gamma\left(\frac{\alpha+2}{2}\right)}
\right)
\cdot
\left(\frac{2k}{q_\mathrm{max}}\right)^\alpha
+O\left(k^2\right)
& , & 0<\alpha<\frac{2}{\pi}\arctan\left(\frac{\mathrm{Re}(m)}{\mathrm{Im}(m)}\right)+1\\
\frac{\mathrm{Im}(m)}{\alpha}
+\frac{\mathrm{Im}(m)}{2(\alpha-2)}
\cdot
\left(\frac{2k}{q_\mathrm{max}}\right)^2
+O\left(k^4\right)
& , & \alpha=\frac{2}{\pi}\arctan\left(\frac{\mathrm{Re}(m)}{\mathrm{Im}(m)}\right)+1\\
\frac{\mathrm{Im}(m)}{\alpha}
+\left(
\mathrm{Im}(m)\cdot\frac{\sqrt{\pi}}{2}\frac{\Gamma\left(-\frac{\alpha}{2}\right)}{\Gamma\left(\frac{1-\alpha}{2}\right)}
-\mathrm{Re}(m)\cdot\frac{\sqrt{\pi}}{2}\frac{\Gamma\left(\frac{\alpha+1}{2}\right)}{\Gamma\left(\frac{\alpha+2}{2}\right)}
\right)
\cdot
\left(\frac{2k}{q_\mathrm{max}}\right)^\alpha
+O\left(k^2\right)
& , & \frac{2}{\pi}\arctan\left(\frac{\mathrm{Re}(m)}{\mathrm{Im}(m)}\right)+1<\alpha<2\\
\frac{\mathrm{Im}(m)}{\alpha}
-\frac{\mathrm{Im}(m)}{2}
\left(\frac{2k}{q_\mathrm{max}}\right)^2
\ln\left(\frac{2k}{q_\mathrm{max}}\right)
+O\left(k^2\right)
& , & \alpha=2\\
\frac{\mathrm{Im}(m)}{\alpha}
+\frac{\mathrm{Im}(m)}{2(\alpha-2)}
\cdot
\left(\frac{2k}{q_\mathrm{max}}\right)^2
+o\left(k^2\right)
& , & \alpha>2
\end{aligned}
\right.
\end{aligned}
\label{c9}
$}
\end{equation}

Then, according to Eq. \eqref{a14}, we get:
\begin{equation}
\scalebox{0.86}{$\displaystyle
\begin{aligned}
\Delta\lambda_\mathbf{k}
=&
-\frac{(\alpha+2)w^2 a^2}{3\cdot\mathrm{Re}(E_\mathbf{k})^{\frac{3}{2}}}\times\\
&\left\{
\begin{aligned}
-\mathrm{Im}(m)\cdot\ln\left(\frac{2k}{q_\mathrm{max}}\right)+O\left(k^0\right) & , & \alpha=0\\
\frac{\mathrm{Im}(m)}{\alpha}
+\left(
\mathrm{Im}(m)\cdot\frac{\sqrt{\pi}}{2}\frac{\Gamma\left(-\frac{\alpha}{2}\right)}{\Gamma\left(\frac{1-\alpha}{2}\right)}
-\mathrm{Re}(m)\cdot\frac{\sqrt{\pi}}{2}\frac{\Gamma\left(\frac{\alpha+1}{2}\right)}{\Gamma\left(\frac{\alpha+2}{2}\right)}
\right)
\cdot
\left(\frac{2k}{q_\mathrm{max}}\right)^\alpha
+O\left(k^2\right)
& , & 0<\alpha<\frac{2}{\pi}\arctan\left(\frac{\mathrm{Re}(m)}{\mathrm{Im}(m)}\right)+1\\
\frac{\mathrm{Im}(m)}{\alpha}
+\frac{\mathrm{Im}(m)}{2(\alpha-2)}
\cdot
\left(\frac{2k}{q_\mathrm{max}}\right)^2
+O\left(k^4\right)
& , & \alpha=\frac{2}{\pi}\arctan\left(\frac{\mathrm{Re}(m)}{\mathrm{Im}(m)}\right)+1\\
\frac{\mathrm{Im}(m)}{\alpha}
+\left(
\mathrm{Im}(m)\cdot\frac{\sqrt{\pi}}{2}\frac{\Gamma\left(-\frac{\alpha}{2}\right)}{\Gamma\left(\frac{1-\alpha}{2}\right)}
-\mathrm{Re}(m)\cdot\frac{\sqrt{\pi}}{2}\frac{\Gamma\left(\frac{\alpha+1}{2}\right)}{\Gamma\left(\frac{\alpha+2}{2}\right)}
\right)
\cdot
\left(\frac{2k}{q_\mathrm{max}}\right)^\alpha
+O\left(k^2\right)
& , & \frac{2}{\pi}\arctan\left(\frac{\mathrm{Re}(m)}{\mathrm{Im}(m)}\right)+1<\alpha<2\\
\frac{\mathrm{Im}(m)}{\alpha}
-\frac{\mathrm{Im}(m)}{2}
\left(\frac{2k}{q_\mathrm{max}}\right)^2
\ln\left(\frac{2k}{q_\mathrm{max}}\right)
+O\left(k^2\right)
& , & \alpha=2\\
\frac{\mathrm{Im}(m)}{\alpha}
+\frac{\mathrm{Im}(m)}{2(\alpha-2)}
\cdot
\left(\frac{2k}{q_\mathrm{max}}\right)^2
+o\left(k^2\right)
& , & \alpha>2
\end{aligned}
\right.
\end{aligned}
\label{c13}
$}
\end{equation}

Here, we use some important facts:
\begin{equation}
\lim_{\alpha\to 0}
\frac{1}{\alpha}
+\frac{\sqrt{\pi}}{2}\frac{\Gamma\left(-\frac{\alpha}{2}\right)}{\Gamma\left(\frac{1-\alpha}{2}\right)}y^\alpha
=-\ln y+O(1),
\label{c10}
\end{equation}
\begin{equation}
\lim_{\alpha\to 2}
\frac{y^2}{2(\alpha-2)}
+\frac{\sqrt{\pi}}{2}\frac{\Gamma\left(-\frac{\alpha}{2}\right)}{\Gamma\left(\frac{1-\alpha}{2}\right)}y^\alpha
=-\frac{1}{2}y^2\ln y+O\left(y^2\right),
\label{c11}
\end{equation}
\begin{equation}
\Gamma\left(x\right)\Gamma\left(1-x\right)
=
\frac{\pi}{\sin\left(\pi x\right)}
\label{c12}
\end{equation}

Equation \eqref{c9} is equivalent to the form we present in Eq. (6) in the main text.

From Eqs. \eqref{c9} and \eqref{c13}, we can see that:

1. When $\alpha=0$, the disorder reduces to an uncorrelated disorder, that the scattering loss $\mathrm{Im}(\Sigma_\mathbf{k})$ and the excess linewidth $\Delta\lambda_\mathbf{k}$ diverge at $k=0$ as $O (\ln k)$.

2. When $\alpha>0$, the leading order in the scattering loss $\mathrm{Im}(\Sigma_\mathbf{k})$ and the excess linewidth $\Delta\lambda_\mathbf{k}$ is no longer $k^\alpha$ (which is the Hermitian case in Eqs. \eqref{b9} and \eqref{b10}), but is always a finite constant (independent of $k$):
\begin{equation}
\left.\mathrm{Im}(\Sigma_\mathbf{k})\right|_{k=0}=\frac{(\alpha+2)w^2 a^2}{6\pi\alpha}\cdot\mathrm{Im}(m),
\label{c14-0}
\end{equation}
\begin{equation}
\left.\Delta\lambda_\mathbf{k}\right|_{k=0}=-\frac{(\alpha+2)w^2 a^2}{3\alpha \cdot \mathrm{Re}(E_\mathbf{k})^{\frac{3}{2}}}\cdot\mathrm{Im}(m).
\label{c14}
\end{equation}

3. When $\alpha=\frac{2}{\pi}\arctan\left(\frac{\mathrm{Re}(m)}{\mathrm{Im}(m)}\right)+1$, there is a sudden jump in the next leading-order scaling, that the second leading order jumps from $O(k^\alpha)$ to $O(k^2)$. This is because the coefficient of the $O(k^\alpha)$ term accidentally reduces to zero when $\alpha=\frac{2}{\pi}\arctan\left(\frac{\mathrm{Re}(m)}{\mathrm{Im}(m)}\right)+1$: 
\begin{equation}
\mathrm{Im}(m)\cdot\frac{\sqrt{\pi}}{2}\frac{\Gamma\left(-\frac{\alpha}{2}\right)}{\Gamma\left(\frac{1-\alpha}{2}\right)}
-\mathrm{Re}(m)\cdot\frac{\sqrt{\pi}}{2}\frac{\Gamma\left(\frac{\alpha+1}{2}\right)}{\Gamma\left(\frac{\alpha+2}{2}\right)}
=
0.
\label{c14-1}
\end{equation}

Hence, the third leading-order term $O(k^2)$ becomes the second leading-order exponent in the excess linewidth.

4. The next leading order (in $k$) of the scattering loss $\mathrm{Im}(\Sigma_\mathbf{k})$ and the excess linewidth $\Delta\lambda_\mathbf{k}$ is:
\begin{equation}
\begin{aligned}
&\mathrm{Im}(\Sigma_\mathbf{k})-\left.\mathrm{Im}(\Sigma_\mathbf{k})\right|_{k=0}
\propto
\Delta\lambda_\mathbf{k}-\left.\Delta\lambda_\mathbf{k}\right|_{k=0}
=
\left\{
\begin{aligned}
O\left(k^\alpha\right)
& , & 0<\alpha<\frac{2}{\pi}\arctan\left(\frac{\mathrm{Re}(m)}{\mathrm{Im}(m)}\right)+1\\
O\left(k^2\right)
& , & \alpha=\frac{2}{\pi}\arctan\left(\frac{\mathrm{Re}(m)}{\mathrm{Im}(m)}\right)+1\\
O\left(k^\alpha\right)
& , & \frac{2}{\pi}\arctan\left(\frac{\mathrm{Re}(m)}{\mathrm{Im}(m)}\right)+1<\alpha<2\\
O\left(k^2\ln k\right)
& , & \alpha=2\\
O\left(k^2\right)
& , & \alpha>2
\end{aligned}
\right.
\end{aligned}
\label{c15}
\end{equation}

This is also very different from the Hermitian case in Eq. \eqref{b10}. Even when $\alpha$ is very large, the next leading order of the scattering loss $\mathrm{Im}(\Sigma_\mathbf{k})$ and the excess linewidth $\Delta\lambda_\mathbf{k}$ can never exceed $O\left(k^2\right)$.

5. When $\mathrm{Im}(m)=0$, Eqs. \eqref{c9} and \eqref{c13} reduce to the Hermitian case in Eqs. \eqref{b9} and \eqref{b10}, respectively.

\end{document}